\documentclass[12pt]{article}
\pdfoutput=1

\usepackage[utf8]{inputenc}
\usepackage[left=2.55cm, right=2.55cm, top=2.55cm, bottom=2.55cm]{geometry}
\usepackage{amsmath,amssymb,amsbsy,mathtools}
\usepackage{slashed}
\usepackage{xcolor}
\usepackage{graphicx}
\usepackage{url}
\usepackage{cancel}
\usepackage{cite}
\usepackage[colorlinks=true,allcolors=blue,pdfborder={0 0 0},linktocpage=false,pdfencoding=auto]{hyperref}
\usepackage{tabularx,booktabs}
\usepackage{multicol}
\usepackage{multirow}
\usepackage{feynmp}
\usepackage{units}
\usepackage{xspace}
\usepackage[labelfont=bf]{caption}
\usepackage[section]{placeins}
\usepackage{subcaption}
\usepackage{soul}
\usepackage{bbold}
\usepackage{parskip}
\usepackage{float}
\usepackage{tabulary}
\usepackage{empheq}
\usepackage{tabu}
\usepackage{listings}

\definecolor{darkred}{rgb}{0.6,0,0}
\definecolor{darkpurple}{rgb}{0.5,0,0.5}
\definecolor{codebg}{RGB}{247,248,250}
\definecolor{codeframe}{RGB}{210,214,220}
\definecolor{codecomment}{RGB}{30,115,70}
\definecolor{codekeyword}{RGB}{40,70,150}
\newcommand{\code}[1]{\texttt{#1}}

\lstdefinestyle{classcode}{
  language=C,
  basicstyle=\ttfamily\small,
  keywordstyle=\color{codekeyword}\bfseries,
  commentstyle=\color{codecomment},
  stringstyle=\color{red!55!black},
  backgroundcolor=\color{codebg},
  frame=single,
  rulecolor=\color{codeframe},
  breaklines=true,
  showstringspaces=false,
  columns=fullflexible,
  keepspaces=true,
  tabsize=2
}

\begin{document}
\author{Amin Aboubrahim$^{a}$\footnote{\href{mailto:abouibrah@hartford.edu}{abouibrah@hartford.edu}}\,~~and Pran Nath$^b$\footnote{\href{mailto:p.nath@northeastern.edu}{p.nath@northeastern.edu}}
 \\~\\
$^{a}$\textit{\normalsize Department of Physics, University of Hartford,} \\
\textit{\normalsize 200 Bloomfield Ave., West Hartford, CT 06117, U.S.A.} \\
$^{b}$\textit{\normalsize Department of Physics, Northeastern University,} \\
\textit{\normalsize 111 Forsyth Street, Boston, MA 02115-5000, U.S.A.} \\
}

\title{\vspace{-2cm}
\vspace{1cm}
\large \bf
Automatic detection of fast oscillations of dark matter scalar field and updated cosmological constraints on QCDM
 \vspace{0.5cm}}
\date{}
\maketitle

\vspace{0.5cm}

\begin{abstract}

It is well known that a scalar field dark matter with a quadratic potential undergoes fast oscillations when the time period represented by the mass scale in the Klein-Gordon equation becomes much smaller than that set by the Hubble parameter. This makes a solution to the equation numerically intractable. Many works in the literature have addressed the problem by either switching between solving the Klein-Gordon equation in the well-behaved regime to solving the fluid equations at the onset of oscillations, or by introducing a new set of variables that can absorb these oscillations. Despite being successful, these techniques rely on an estimate of when the oscillations start. For large scale scans of a model's parameter space, this can become cumbersome. Furthermore, the techniques have been used mainly for non-interacting dark matter models. In this work, we introduce an averaging technique with an automatic detection of the onset of oscillations, capable of capturing the non-interacting as well as the interacting dark matter scenarios. The technique, implemented in \code{CLASS}, is tested on the QCDM model and shows excellent detection and averaging abilities. We also update the cosmological constraints on the model using the most recent public data. 

\end{abstract}

\numberwithin{equation}{section}

\newpage

{  \hrule height 0.4mm \hypersetup{linktocpage=true} \tableofcontents
\vspace{0.5cm}
 \hrule height 0.4mm}

\section{Introduction}

Ultralight spin-zero fields show a remarkable success in capturing the behavior of the dark sector elements of cosmology, i.e., dark matter and dark energy. A minimally coupled spin-zero field can exhibit a rich phenomenology over the history of cosmic evolution depending on the form of its potential and the cosmic epoch under consideration. This versatility has motivated the use of these fields as candidates for both dark energy and dark matter, thus replacing the cold dark matter (CDM) paradigm in concordance $\Lambda$CDM model~\cite{Hu:2000ke,Ringwald:2012hr,Hui:2016ltb}. A dark matter scalar field undergoes rapid oscillations around the minimum of its potential after the field overcomes the Hubble friction. These coherent oscillations give an energy density that dilutes with the expansion of the universe as matter does, i.e., the scalar field mimics a pressureless fluid. On the other hand, a slowly-varying field can generate negative pressure, thus driving the late-time accelerated expansion of the universe. A popular example of such a field is known as quintessence~\cite{Ratra:1987rm,Caldwell:1997ii,Wetterich:1987fm,Albrecht:1999rm,Wetterich:2004pv} (for a review see ref.~\cite{Tsujikawa:2013fta} and references therein). Unlike the cosmological constant, quintessence possesses a dynamical background evolution and a non-trivial perturbation which can leave observable signatures in the expansion history and cosmological structure. 

A scalar field dark matter with a quadratic potential~\cite{Sahni:1999qe,Matos:2000ng,Amendola:2005ad,Matos:2008ag,Hwang:2009js,Marsh:2010wq,Glennon:2023jsp,Park:2012ru,Li:2013nal,Urena-Lopez:2015gur} (see refs.~\cite{Marsh:2015xka,Magana:2012xe} for reviews) behaves as pressureless matter at the onset of rapid oscillations, which happens when the Hubble rate, $H$, becomes less than the mass parameter in the potential. This has been realized and studied in ultralight axion and fuzzy dark matter scenarios~\cite{Matos:2000ss,Hu:2000ke,Marsh:2010wq}, where for a potential of the form $V(\chi)\propto |\chi|^n$, the oscillation-averaged energy density redshifts as
\begin{equation}
\langle\rho_\chi\rangle\propto a^{-3(1+\langle w_\chi\rangle)},
\end{equation}
where $a$ is the scale factor and the effective equation of state $\langle w_\chi\rangle=\langle p_\chi\rangle/\langle\rho_\chi\rangle$ is
\begin{equation}
\langle w_\chi\rangle=\frac{n-2}{n+2}.
\end{equation}
For the quadratic case, $n=2$, $\langle w_\chi\rangle=0$ and so the rapidly oscillating scalar field behaves like nonrelativistic matter. 
The inclusion of anharmonic corrections resulting from the addition of the self-interaction quartic term~\cite{Cembranos:2018ulm,Foidl:2022bpn} has been shown to alter the pressure of the field which leads to experimental constraints on the self-interaction strength~\cite{Aboubrahim:2024spa}. Linear perturbations in scalar fields introduce a characteristic scale below which a pressure perturbation can sustain a non-negligible sound speed which causes a scale-dependent suppression of the matter transfer function. This is known to happen in the case of ultralight scalar fields and fuzzy dark matter. For scalar fields with a quadratic potential (and mass $m_\chi$), the Jeans scale 
\begin{equation}
    k_J\sim a\sqrt{m_\chi H}\,,
\end{equation}
tells us that for sufficiently large dark matter masses, the scalar field will reproduce the exact CDM behavior while smaller masses will produce observable oscillatory features suppressing small-scale linear power~\cite{Hu:2000ke,Urena-Lopez:2015gur}. 

Even though the unique and characteristic feature of rapid oscillations of scalar fields is important to reproduce the CDM behavior, it presents a challenge when numerically solving for the evolution of the field. When the mass scale in the problem becomes larger than the Hubble parameter, an increasingly large number of oscillations must be resolved during each cosmological timescale. This becomes particularly problematic especially in Boltzmann solvers like \code{CLASS}~\cite{Blas:2011rf} where the background and perturbations modules are run repeatedly to evaluate the cosmology. The problem is made worse when a Markov Chain Monte Carlo simulator is involved since these calculations may be repeated thousands of times. The treatment of the transition from slow field variations to the rapid oscillations is therefore an essential component of precision scalar field dark matter cosmology~\cite{Lesgourgues:2011re,Urena-Lopez:2015gur}. A common strategy in the literature is to solve for the exact evolution of the slowly-varying field prior to oscillations and then transition to solving the averaged fluid equations at the onset of oscillations, which has shown to be very useful in studying the CMB and matter power spectra of ultralight axions\footnote{This strategy has been implemented in a numerical code called \code{axionCAMB}~\cite{Hlozek:2014lca}.}. The accuracy of this method has been investigated extensively. For example, the authors of ref.~\cite{Cookmeyer:2019rna} have demonstrated that inaccuracies in the averaging methods can propagate into a number of precision cosmological observables, thus biasing the outcome. More recently, ref.~\cite{Passaglia:2022bcr} came up with an improved effective-fluid description designed to reproduce the exact axion evolution to the sub-percent accuracy. An alternative approach was introduced by refs.~\cite{Urena-Lopez:2015gur,Urena-Lopez:2023ngt}, where new variables, including a polar angle $\theta$, were used to turn the Klein-Gordon equation into a set of coupled first order differential equations. Interestingly, the equation of state of the scalar field can be written as $w_\chi=-\cos\theta$, which makes the averaging process more straightforward. Detailed numerical studies have shown that such formulations can reproduce the original scalar-field equations accurately while substantially improving their practical treatment in a Boltzmann code  

All of the above techniques discussed in the literature rely on a criterion to tell us when these rapid oscillations in the scalar field start, and it usually pertains to the smallness of the ratio $H/m_\chi$. Once this ratio hits a small enough value, the averaging is switched on. If the chosen criterion forces an early transition, then the averaging scheme will be implemented prematurely, whereas a late transition will unnecessarily force the integrator to resolve the rapid oscillations causing a slow down of the process. It may be straightforward to predict the transition scale for simple models, but this quickly becomes cumbersome in multi-field models with interactions involved and where large parts of the parameter space need to be explored. The accuracy of the handoff between exact and averaged evolution is an important ingredient in precision predictions~\cite{Cookmeyer:2019rna,Passaglia:2022bcr}.  

In this work, we present a technique to automatically detect the time of the onset of oscillations of a scalar field and implement it in \code{CLASS}. Rather than requiring the user to guess the criterion at which averaging needs to be activated, the algorithm automatically detects a specific change in the evolution of some dynamical variables which signals the start of the oscillatory phase. The code then employs an independent method to confirm that the measured changes indeed correspond to the beginning of these rapid oscillations. After the transition point is detected, the averaging procedure is triggered. The algorithm therefore determines the transition from the field dynamics itself, while retaining a manual mode for direct numerical comparison and validation. We apply this method of automatic detection and averaging to a field-theoretic model of cosmology, known as QCDM~\cite{Aboubrahim:2025usl,Aboubrahim:2024spa,Aboubrahim:2024cyk,Aboubrahim:2026tks,Nath:2025gzm}. We then use the most up-to-date cosmological data to derive constraints on the cosmological parameters of the model and compare the fits to those of $\Lambda$CDM. 

The rest of the paper is organized as follows: in section~\ref{sec:simple} we go over a simple model which uses an averaging technique that is common in the literature, and in section~\ref{sec:interacting} the QCDM model is introduced as a field-theoretic model of interacting dark matter and dark energy. We study the effect of this interaction on the dark matter equation of state in section~\ref{sec:eos} and explain why the standard averaging technique can fail in this situation. In sections~\ref{sec:averaging} and~\ref{sec:automatic} we introduce the new averaging technique that addresses the presence of interactions and the automatic detection method of the onset of oscillations and describe the implementation in \code{CLASS}. We explain in section~\ref{sec:perturb} the effect of the interaction on perturbations and then extract the new constraints on the model parameters in section~\ref{sec:constraints}. We end in section~\ref{sec:conclusion} with conclusions.

\section{The simple model}\label{sec:simple}

A minimal modification to $\Lambda$CDM involves replacing a fluid-like cold dark matter with an ultralight scalar field $\chi$ having a simple quadratic potential $V(\chi)=(1/2)m_\chi^2\chi^2$. The energy density and pressure of the field are
\begin{align}
    \rho_\chi&=\frac{1}{2a^2}\chi^{\prime 2}+V(\chi)\,, \\
    p_\chi&=\frac{1}{2a^2}\chi^{\prime 2}-V(\chi)\,,
\end{align}
where $a$ is the scale factor and a prime represents derivative with respect to conformal time. The DM field obeys the Klein-Gordon (KG) equation
\begin{equation}
    \chi^{\prime\prime}+2\mathcal{H}\chi^\prime+a^2 m_\chi^2 \chi=0\,,
\end{equation}
where $\mathcal{H}=a^\prime/a$ is the conformal Hubble parameter. In this equation, there are two competing time scales, $m_\chi^{-1}$ and $\mathcal{H}^{-1}$. Given that the DM field is ultralight, there will be a time during the cosmic history when $\mathcal{H}/(a\,m_\chi)\ll 1$ which causes the DM field $\chi$ to oscillate violently at the minimum of its potential. As a consequence, the field will start to behave as cold dark matter with an energy density that dilutes as $\rho_\chi\sim a^{-3}$. However, any numerical code will face difficulty in this regime since the integrator will take progressively smaller steps to try and overcome the stiffness of the KG equation, causing the code to stall. 

To overcome this difficulty, one can solve the KG equation in the well-behaved regime and then switch to solving the fluid equations once oscillations start. This requires a criterion that triggers the code to perform this switch, and in general is chosen to be when $\sqrt{\partial^2V/\partial\chi^2}>3H$~\cite{Cembranos:2018ulm}, where $H=\dot{a}/a$ is the Hubble parameter in cosmic time. This choice is arbitrary and can fail in more complicated models and when exploring larger parts of the parameter space. Another method would be to introduce a new set of variables and convert the KG equation into a set of coupled first order differential equations which can be easily handled. We now briefly go over the latter method which has previously been discussed in the literature.

Thus, we introduce three dimensionless variables, $\Omega_\chi$, $\theta$ and $y$, defined as~\cite{Copeland:1997et,Garcia-Arroyo:2024tqq}
\begin{align}
&{\Omega}_\chi^{1/2}\sin\left(\frac{\theta}{2}\right)=\frac{\kappa \chi^\prime}{\sqrt{6}\mathcal{H}}, \qquad {\Omega}_\chi^{1/2}\cos\left(\frac{\theta}{2}\right)=\frac{\kappa a V^{1/2}}{\sqrt{3}\mathcal{H}}, \qquad y=-\frac{2\sqrt{2}\,a}{\mathcal{H}}\partial_\chi V^{1/2}\,,
\end{align}
where $\kappa\equiv \sqrt{8\pi G}$ and $\Omega_\chi=\kappa^2 a^2\rho_\chi/3\mathcal{H}^2$ is the usual energy density fraction. One can then show that the KG equation of $\chi$ can be turned into the following differential equations
\begin{align}
\label{om0}
&\Omega^\prime_\chi=3\mathcal{H}\Omega_\chi(w_T-w_\chi)\,, \\
\label{theta0}
&\theta^\prime=-3\mathcal{H}\sin\theta+\mathcal{H}y\,, \\
&y^\prime=\frac{3}{2}{\cal H}(1+w_T)y\,,
\label{y0}
\end{align}
where $w_T=\sum p_i/\sum\rho_i$ is the total equation of state, with the sum running over all species (baryons, photons, neutrinos, DM and DE). One can easily show that the equation of state of $\chi$, conveniently, becomes
\begin{equation}
    w_\chi\equiv \frac{p_\chi}{\rho_\chi}=-\cos\theta\,.
\end{equation}
During the rapid oscillations that the field undergoes, the equation of state oscillates between $-1$ and $+1$ and the fact that $w_\chi=-\cos\theta$ makes the averaging process straightforward, i.e., $\langle w_\chi\rangle=-\langle\cos\theta\rangle=0$, as intended. This procedure usually implements a cutoff scale which is introduced by defining~\cite{Urena-Lopez:2015gur}
\begin{equation}
    \{\cos_*\gamma\,,\sin_*\gamma\}\equiv \frac{1}{2}\left[1-\tanh(\gamma^2-\gamma_*^2)\right]\{\cos\gamma\,,\sin\gamma\}\,.
    \label{star}
\end{equation}
In this case, for $\gamma<\gamma_*$, $\{\cos_*\gamma\,,\sin_*\gamma\}\to \{\cos\gamma\,,\sin\gamma\}$, while for $\gamma>\gamma_*$, $\{\cos_*\gamma\,,\sin_*\gamma\}\to 0$. The cutoff scale $\gamma_*$ is guessed by trial and error and a convenient value for the non-interacting case turns out to be $\gamma_*\simeq 100$. However, this value may fluctuate and can greatly differ for very different models, especially in the presence of interactions.

\section{The interacting model}\label{sec:interacting}

In a field-theoretic analogue to the standard model of cosmology, we introduce two interacting ultralight spin-zero fields: one is a scalar dark matter (DM) field, $\chi$, and the other is a quintessence-type pseudoscalar, $\phi$. We call this model QCDM, with an action given by~\cite{Aboubrahim:2025usl,Aboubrahim:2024spa,Aboubrahim:2024cyk,Aboubrahim:2026tks,Nath:2025gzm}
\begin{align}
 &S_{\rm QCDM}=\int \text{d}^4 x\,\sqrt{-g}
  \left[\frac{1}{16\pi G}R+\frac{1}{2}\partial_\mu \phi\partial^\mu\phi+\frac{1}{2}\partial_\mu \chi\partial^\mu\chi-
  V(\phi,\chi)\right],
\end{align}
where the total potential is
\begin{align}
  &V(\phi,\chi)=V_1(\chi)+V_2(\phi)+ V_{\rm int}(\phi,\chi)\,.
\end{align}
For this model, we consider the following forms for the DM, DE and interaction potentials:
\begin{align}
    \label{v1}
    V_1(\chi)&=\frac{1}{2}m_\chi^2\chi^2\,, \\
    V_2(\phi)&=\mu^4\left[1+\cos\left(\frac{\phi}{F}\right)\right]\,, \\
    V_{\rm int}(\phi,\chi)&=\frac{\lambda}{2}\chi^2\phi^2\,,
    \label{vint}
\end{align}
where $m_\chi$ is the DM bare mass, $\mu$ and $F$ are constants and $\lambda$ represents the DM-DE interaction strength. 

The perturbed spacetime FLRW metric in the synchronous gauge is 
\begin{equation}
    \text{d}s^2=a(\tau)^2\big[-\text{d}\tau^2+(\gamma_{ij}+h_{ij})\text{d}x^i\text{d}x^j \big]\,,
\end{equation}
where $\tau$ is the conformal time, $\gamma_{ij}=\delta_{ij}$ for flat spacetime and $h_{ij}$ is a spatial perturbation. The DM and DE fields are also perturbed as well as the stress-energy tensor, so that
\begin{align}
    \chi(\tau,\mathbf{x})&= \chi_0(\tau)+\chi_1(\tau,\mathbf{x})\,, \\
    \phi(\tau,\mathbf{x})&= \phi_0(\tau)+\phi_1(\tau,\mathbf{x})\,, \\
    T^{\mu\nu}(\tau,\mathbf{x})&=\bar{T}^{\mu\nu}(\tau)+\delta T^{\mu\nu}(\tau,\mathbf{x})\,, 
\end{align}
where the elements of the mixed stress-energy tensor are 
\begin{align}
    T^{0}_0&=-\rho-\delta\rho\,, \nonumber \\
    T^{0}_i&=(\rho+p)v_i\,, \nonumber \\
    T^{i}_0&=-(\rho+p)v_i\,, \nonumber \\
    T^{i}_j&= (p+\delta p)\delta^i_j+p\Pi^i_j \,.
\end{align}
Here $\Pi^i_j$ represents the anisotropic stress, $v_i$ the 3-velocity and  $\delta\rho$ and $\delta p$ are the density and pressure perturbations, respectively.

As a result, the energy density and pressure of the background fields are
\begin{align}
    Z^{\pm}_\chi=\frac{1}{2a^2}\chi_0^{\prime 2}\pm V_1(\chi_0)\pm V_{\rm int}(\phi_0,\chi_0)\,, \\
    Z^{\pm}_\phi=\frac{1}{2a^2}\phi_0^{\prime 2}\pm V_2(\phi_0)\pm V_{\rm int}(\phi_0,\chi_0)\,, 
\end{align}
where $Z^+_i$ represents the energy density $\rho_i$ and $Z_i^-$ the pressure $p_i$.  
The corresponding continuity equations for the background fields then read
\begin{align*}
&\rho^\prime_\phi+3\mathcal{H}(1+w_\phi)\rho_\phi=Q_\phi\equiv \chi^\prime_0 V_{\rm int,\chi} \,, \\
&\rho^\prime_\chi+3\mathcal{H}(1+w_\chi)\rho_\chi=Q_\chi\equiv \phi^\prime_0 V_{\rm int,\phi}\,,
\end{align*}
where the notation $V_{,Y}$ means $\partial V/\partial Y$, $\mathcal{H}=a^\prime/a$ is the conformal Hubble parameter and $w=p/\rho$ is the equation of state. To make sure conservation of energy of the \{DM, DE\} system while preserving diffeomorphism of the theory, we define the total energy density as
\begin{equation}
\rho=\rho_\phi+\rho_\chi-V_{\rm int}(\phi,\chi)\,,
\end{equation}
which prevents a double-counting of the interaction potential and satisfies $\rho^\prime+3\mathcal{H}(\rho+p)=0$. The evolutions of the background fields $\chi_0$ and $\phi_0$ are determined by solving the Klein-Gordon (KG) equations
\begin{align}
\label{kgc0}
&\chi_0^{\prime\prime}+2\mathcal{H}\chi_0^\prime+a^2(V_{1,\chi}+V_{\text{int},\chi})=0, \\
&\phi_0^{\prime\prime}+2\mathcal{H}\phi_0^\prime+a^2(V_{2,\phi}+V_{\text{int},\phi})=0\,.
\label{kgp0}
\end{align}
The density and pressure perturbations of the two fields, $\chi$ and $\phi$, in the synchronous gauge are given by
\begin{align}
    \label{dzp}
    \delta Z^{\pm}_\phi&=\frac{1}{a^2}\phi_0^\prime\phi_1^\prime\pm ({V}_2+{V}_{\rm int})_{,\phi}\phi_1\pm {V}_{\text{int},\chi}\chi_1\,, \\
    \delta Z^{\pm}_\chi&=\frac{1}{a^2}\chi_0^\prime\chi_1^\prime\pm ({V}_1+{V}_{\rm int})_{,\chi}\chi_1\pm {V}_{\text{int},\phi}\phi_1\,,
    \label{dzm}
\end{align} 
where, again, $\delta Z_i^+\equiv \delta\rho_i$ and $\delta Z_i^-\equiv \delta p_i$. In terms of the wavenumber $k$, the momentum densities of the fields are
\begin{align}
    (\rho_\phi+p_\phi)\Theta_\phi&=\frac{k^2}{a^2}\phi_0^\prime\phi_1, \\
    \label{divchi}
    (\rho_\chi+p_\chi)\Theta_\chi&=\frac{k^2}{a^2}\chi_0^\prime\chi_1\,,
\end{align}
where $\Theta=ik^i v_i$ is the velocity divergence. Determining the perturbations in Eqs.~(\ref{dzp}) and~(\ref{dzm}) requires solving the KG equations for the background, Eqs.~(\ref{kgc0}) and~(\ref{kgp0}), and the KG equations for the perturbations. The latter are given by
\begin{align}
\label{KGp1}
&\phi_1^{\prime\prime}+2\mathcal{H}\phi_1^\prime+(k^2+a^2 V_{,\phi\phi})\phi_1+a^2{V}_{,\phi\chi}\chi_1+\frac{1}{2}h^\prime\phi_0^\prime=0, \\
&\chi_1^{\prime\prime}+2\mathcal{H}\chi_1^\prime+(k^2+a^2{V}_{,\chi\chi})\chi_1+a^2{V}_{,\chi\phi}\phi_1+\frac{1}{2}h^\prime\chi_0^\prime=0,
\end{align}
where $V_{,XX}=\partial^2 V/\partial X^2$ and $V_{,XY}=\partial^2 V/\partial X\partial Y$.

We would like to make use of the new variables $\Omega$, $\theta$ and $y$ introduced earlier. However, for the interaction case, it is easier to deal with the modified energy density $\tilde\rho_\chi=\rho_\chi-V_{\rm int}$, such that $\tilde\Omega_\chi=\kappa^2 a^2\tilde\rho_\chi/3\mathcal{H}^2$. We can then add the interaction term in the final equations of the analysis. Therefore, the new background variables become
\begin{align}
    \label{var-tilde-1}
    &\tilde{\Omega}_\chi^{1/2}\sin\left(\frac{\theta}{2}\right)=\frac{\kappa \chi^\prime}{\sqrt{6}\mathcal{H}}, \\
    \label{var-tilde-2}
    &\tilde{\Omega}_\chi^{1/2}\cos\left(\frac{\theta}{2}\right)=\frac{\kappa a V^{1/2}_{1}}{\sqrt{3}\mathcal{H}}, \\
    &y=-\frac{2\sqrt{2}\,a}{\mathcal{H}}\partial_\chi V^{1/2}_{1}.
    \label{var-tilde-3}
\end{align}
When it comes to the perturbations, we define the DM density perturbation by absorbing the interaction term into the $\delta\tilde{\rho}_\chi$, so that
\begin{equation}
   \delta\tilde{\rho}_\chi=\delta\rho_\chi-V_{\text{int},\phi}\phi_1-V_{\text{int},\chi}\chi_1 \,.
   \label{dtr}
\end{equation} 
In this notation, the density contrast is $\tilde\delta_\chi=\delta\tilde\rho_\chi/\tilde\rho_\chi$. A new set of dimensionless variables can be introduced as well for the perturbations  
\begin{align}
\label{cvt}
&\sqrt{\frac{2}{3}}\frac{\kappa\,\chi^\prime_1}{\mathcal{H}}=-\tilde{\Omega}_\chi^{1/2}e^\alpha\cos\left(\frac{\vartheta}{2}\right), \\
\label{svt}
&\frac{\kappa\,y\,\chi_1}{\sqrt{6}}=-\tilde{\Omega}_\chi^{1/2}e^\alpha\sin\left(\frac{\vartheta}{2}\right), \\
\label{tdc}
&\tilde{\delta}_\chi=-e^\alpha\sin\left(\frac{\theta-\vartheta}{2}\right), \\
&\tilde{\Delta}_\chi=-e^\alpha\cos\left(\frac{\theta-\vartheta}{2}\right)\,,
\label{tcdc}
\end{align}
where $e^\alpha$ represents the amplitude of perturbations, a placeholder that will ultimately drop out of the calculations. The quantity $\tilde\Delta_\chi$ is related to the pressure perturbation $\delta\tilde p_\chi$ so that
\begin{equation}
\delta\tilde p_\chi=\tilde\rho_\chi (\tilde\Delta_\chi\sin\theta-\tilde\delta_\chi\cos\theta).
\label{dtp}
\end{equation}
These variables allow us to determine the background and perturbation evolution equations which were derived in an earlier work~\cite{Aboubrahim:2025usl}. 

Once the energy density fraction $\Omega_\chi$ is determined, the energy density of the DM field is found using
\begin{equation}
    \rho_\chi=\frac{\Omega_\chi}{1-\Omega_\chi}\sum_{i\neq\chi}\rho_i\,,
\end{equation}
which is a more convenient way to determine the energy density in \code{CLASS}. Furthermore, the determination of $\tilde\delta_\chi$ allows for the calculation of $\delta\tilde\rho_\chi$ followed by $\delta\rho_\chi$ using Eq.~(\ref{dtr}). Similarly, after determining the pressure perturbation from Eq.~(\ref{dtp}), we deduce $\delta p_\chi$ using
\begin{equation}
   \delta\tilde{p}_\chi=\delta p_\chi+V_{\text{int},\phi}\phi_1+V_{\text{int},\chi}\chi_1 \,.
   \label{tpc}
\end{equation} 
This set of equations constitute a closed system for determining the background and perturbation parameters. However, the prescription for averaging introduced in Eq.~(\ref{star}) does not capture the presence of the interaction term. When applied, it leaves a residual pressure component that contributes to the DM's equation of state during rapid oscillations. In the next section, we explain this subtle issue.

\section{The effect of DM-DE interaction on the equation of state}\label{sec:eos}

The KG equation of $\chi$, Eq.~(\ref{kgc0}), can be cast in the form
\begin{equation}
    \chi_0^{\prime\prime}+2\mathcal{H}\chi_0^\prime+a^2 m^2_{\rm eff}\chi_0=0\,,
\end{equation}
where we defined the effective mass as $m_{\rm eff}^2(\tau)=m_\chi^2+\lambda\phi^2$ which includes the interaction strength and the DE field. During the period of rapid oscillations of $\chi$, the DE field $\phi$ is still rolling gently down its potential which makes $m_{\rm eff}(\tau)$ a slowly-varying function of time. Next, we define $u=a(\tau)\chi_0$ so that the KG equation takes the form
\begin{equation}
    u^{\prime\prime}+a^2m_{\rm eff}^2\left(1-\frac{\mathcal{H}^2}{a^2 m_{\rm eff}^2}-\frac{\mathcal{H}^\prime}{a^2 m_{\rm eff}^2}\right)u=0\,.
\end{equation}
Recall that during fast oscillations, $\varepsilon=\mathcal{H}/(a\,m_{\rm eff})\ll 1$, and so the ratios in the parenthesis can be ignored. Using the WKB approximation~\cite{Cembranos:2015oya}, we get
\begin{equation}
    \chi(\tau)\simeq \frac{\chi_c}{a^{3/2}\sqrt{m_{\rm eff}(\tau)}}\cos\left(\int^\tau m_{\rm eff}(\tau^\prime)a(\tau^\prime)\,\text{d}\tau^\prime\right)\,,
\end{equation}
where $\chi_c$ is a constant of integration. Setting $\chi_{\rm m}\equiv \chi_c/a^{3/2}\sqrt{m_{\rm eff}(\tau)}$, we find
\begin{equation}
    \chi^\prime(\tau)=-a\chi_{\rm m}m_{\rm eff}(\tau)\sin\left(\int^\tau m_{\rm eff}(\tau^\prime)a(\tau^\prime)\,\text{d}\tau^\prime\right)+\mathcal{O}\left(\mathcal{H}\chi_{\rm m},\chi_{\rm m}\frac{m_{\rm eff}^\prime}{m_{\rm eff}}\right)\,.
\end{equation}
We can then easily find the corresponding averages
\begin{align}
    \label{cav}
    \langle\chi^2\rangle&=\frac{\chi^2_{\rm m}}{2}\,, \\
    \langle\chi^{\prime 2}\rangle&=\frac{\chi_{\rm m}^2a^2 m_{\rm eff}^2}{2}=a^2m_{\rm eff}^2\langle\chi^2\rangle\,.
    \label{cpav}
\end{align}
It follows that the average of the energy density is $\langle\rho_\chi\rangle=m_{\rm eff}^2\langle\chi^2\rangle$ and so we write
\begin{equation}
    \langle\chi^2\rangle=\frac{\langle\rho_\chi\rangle}{m_{\rm eff}^2(\tau)}\,.
\end{equation}
Taking the average of the interaction potential yields
\begin{equation}
    \langle V_{\rm int}\rangle=\frac{\lambda\phi^2}{2m_{\rm eff}^2}\langle\rho_\chi\rangle\equiv c\langle\rho_\chi\rangle\,.
\end{equation}
Then, the definition $\tilde\rho_\chi=\rho_\chi-V_{\rm int}$ leads to
\begin{equation}
    \langle\tilde\rho_\chi\rangle=(1-c)\langle\rho_\chi\rangle\,.
\end{equation}
Using Eqs.~(\ref{cav}) and~(\ref{cpav}), we find the average of the pressure to be
\begin{equation}
    \langle p_\chi\rangle=\frac{1}{2}\left(\frac{1}{a^2}\langle\chi^{\prime 2}\rangle-m_{\rm eff}^2\langle\chi^2\rangle\right)=0\,.
    \label{zerop}
\end{equation}
It follows that the contribution of the DM-DE interaction to the averaged equation of state must be zero. Based on our definition $\tilde p_\chi=p_\chi+V_{\rm int}$ and taking the average we get the relations
\begin{equation}
    \langle\tilde p_\chi\rangle=\langle V_{\rm int}\rangle=c\langle\rho_\chi\rangle\,, \quad \langle p_\chi\rangle=0\,.
\end{equation}
If one naively implements the averaging based on Eq.~(\ref{star}), we get $\langle\tilde p_\chi\rangle=0$, which means that $\langle p_\chi\rangle=-\langle V_{\rm int}\rangle$. Therefore, the equation of state $w_\chi=\langle p_\chi\rangle/\langle\rho_\chi\rangle$ will erroneously pick up an extra contribution from the interaction term in contradiction with the conclusion from Eq.~(\ref{zerop}). This shows that a new averaging approach must be implemented numerically to avoid this problem.

\section{Averaging in the presence of interactions}\label{sec:averaging}

In this section we present the modified equations that allow for a consistent treatment of averaged quantities in the presence of interactions. We first define the transition function $T(\theta)$
\begin{equation}
    T(\theta)=\frac{1}{2}\left[1-\tanh\left(\frac{\theta-\theta_*}{w}\right)\right]\,,
    \label{starw}
\end{equation}
where $\theta_*$ represents the value of $\theta$ where the transition into the oscillatory regime happens and $w$ is the width of the transition. We will describe in the next section how the \code{CLASS} implementation can automatically detect $\theta_*$ without the need for it to be specified by the user. We define the following quantities based on $T(\theta)$
\begin{align}
    \mathcal{V}_{\rm int}&=TV_{\rm int}+(1-T)\langle V_{\rm int}\rangle\,, \\
    \tilde\rho_\chi&=\rho_\chi-\mathcal{V}_{\rm int}\,, \\
    \tilde p_\chi&=-T\tilde\rho_\chi\cos\theta+(1-T)\langle V_{\rm int}\rangle\,, \\
    p_\chi&=\tilde p_\chi-\mathcal{V}_{\rm int}\,,
\end{align}
where $\mathcal{V}_{\rm int}$ now represents the interaction term.
In the exact regime, $T\to 1$ while at the onset of oscillations $T\to 0$. From the above equations, we can clearly see that when $T\to 0$, $p_\chi\to 0$ as required. The square of the field $\chi$ is also replaced by the definition
\begin{equation}
    \mathcal{X}^2=T\chi^2+(1-T)\langle\chi^2\rangle\,.
\end{equation}
Therefore, the DM potential and the interaction potential are replaced
\begin{align}
    V_1(\chi)&\longrightarrow \mathcal{V}_1(\mathcal{X})=\frac{1}{2}m_\chi^2\mathcal{X}^2\,, \\
    V_{\rm int}&\longrightarrow \mathcal{V}_{\rm int}=\frac{\lambda}{2}\phi^2\mathcal{X}^2\,,
\end{align}
and a derivative with respect to $\chi$ becomes a derivative with respect to $\mathcal{X}$. Trigonometric factors are weighted similarly to Eq.~(\ref{star}) so that $\cos\theta\to T\cos\theta$ and $\sin\theta\to T\sin\theta$ (and the same applies for half-angle expressions). Furthermore, there are combinations of the form $\chi\cos(\theta/2)$ that appear as a result of the interaction term and these are replaced by
\begin{align}
\label{chicos}
\chi\cos\frac{\theta}{2}&\longrightarrow T\chi\cos\frac{\theta}{2}+(1-T)\sqrt{\frac{\mathcal{X}^2}{2}}\,, \\
\chi\sin\frac{\theta}{2}&\longrightarrow T\chi\sin\frac{\theta}{2}\,.    
\label{chisin}
\end{align}
With this prescription described thus far, we made sure that the averaged equation of state of DM does not receive any contribution from the interaction term. We modified the equations to ensure a consistent analysis for the background equations which allows us to safely use the new variables introduced in Eqs.~(\ref{var-tilde-1})$-$(\ref{var-tilde-3}). 

A similar approach must be implemented for the perturbations as well. For the density perturbations, we write Eq.~(\ref{dtr}) as
\begin{equation}
    \delta\tilde\rho_\chi=\delta\rho_\chi-\lambda\phi\phi_1\chi^2-\lambda\phi^2\chi\chi_1\equiv \delta\rho_\chi-\delta V_{\rm int}\,,
\end{equation}
where we introduced the quantity $\delta V_{\rm int}=\lambda\phi\phi_1\chi^2+\lambda\phi^2\chi\chi_1$. Using Eqs.~(\ref{svt}),~(\ref{tdc}) and~(\ref{tcdc}), we can determine the expression of $\chi\chi_1$ to be
\begin{equation}
    \chi\chi_1=-\frac{\sqrt{6\tilde\Omega_\chi}}{\kappa y}\left(\tilde\delta_\chi \chi\cos\frac{\theta}{2}-\tilde\Delta_\chi \chi\sin\frac{\theta}{2}\right)\,,
\end{equation}
with the replacement of $\chi\cos(\theta/2)$ and $\chi\sin(\theta/2)$ understood based on Eqs.~(\ref{chicos}) and~(\ref{chisin}). 

In the fast oscillations regime, the DM scalar field behaves as cold dark matter and so $\delta p_\chi$ must be zero. Similar to the background case where we made sure $\langle p_\chi\rangle=0$ in the oscillations regime, we need to modify the pressure perturbation equations to give us a consistent $\delta p_\chi$ value in the limit $T\to 0$. Note that Eq.~(\ref{tpc}) is simply $\delta\tilde p_\chi=\delta p_\chi+\delta V_{\rm int}$. So, we define
\begin{align}
    \delta\mathcal{V}_{\rm int}&=T\delta V_{\rm int}+(1-T)\langle\delta V_{\rm int}\rangle\,, \\
    \delta\tilde{p}_\chi&=T\tilde\rho_\chi (\tilde\Delta_\chi\sin\theta-\tilde\delta_\chi\cos\theta)+(1-T)\Big(\langle\delta V_{\rm int}\rangle+c_{s,\chi}^2\delta\bar\rho_\chi\Big)\,, \\
    \delta p_\chi&=\delta\tilde{p}_\chi-\delta\mathcal{V}_{\rm int}\,,
\end{align}
where $c_{s,\chi}^2$ is the DM sound speed in the rest frame
\begin{equation}
    c_{s,\chi}^2=\frac{k^2}{k^2+4a^2 m_{\rm eff}^2}\,, \quad m_{\rm eff}^2=m_\chi^2+\lambda\phi^2\,,
\end{equation}
and the comoving (rest frame) density perturbation is given by~\cite{Hu:2003hjx}
\begin{equation}
    \delta\bar\rho_\chi=\delta\rho_\chi-\rho^\prime_\chi\frac{\theta_\chi}{k^2}\,.
\end{equation}
It is easy to see that when $T\to 0$, we get
\begin{equation}
    \langle\delta\tilde p_\chi\rangle\to \langle\delta V_{\rm int}\rangle+c_{s,\chi}^2\delta\bar\rho_\chi\,,\quad \langle\delta p_\chi\rangle\to c_{s,\chi}^2\delta\bar\rho_\chi\,.
\end{equation}
So, now we have a complete set of consistent background and perturbation equations that capture the proper averaging when an interaction term is present.

\section{Automatic detection of the transition phase}\label{sec:automatic}

In Eq.~(\ref{starw}), we introduced the phase $\theta_*$ which represents the value of $\theta$ when fast oscillations of the DM field begin. As discussed before, this value is usually chosen by running multiple points over a range of values and inspecting when oscillations start. This is of course not ideal as the user-defined value can change depending on the parts of the parameter space being investigated and on whether an interaction term is present or not. We discuss here a method that can be implemented in \code{CLASS} to allow the code to automatically detect the beginning of the oscillatory period and commence with the averaging procedure. 

\begin{figure}[t]
\begin{centering}
\includegraphics[width=1.0\linewidth]{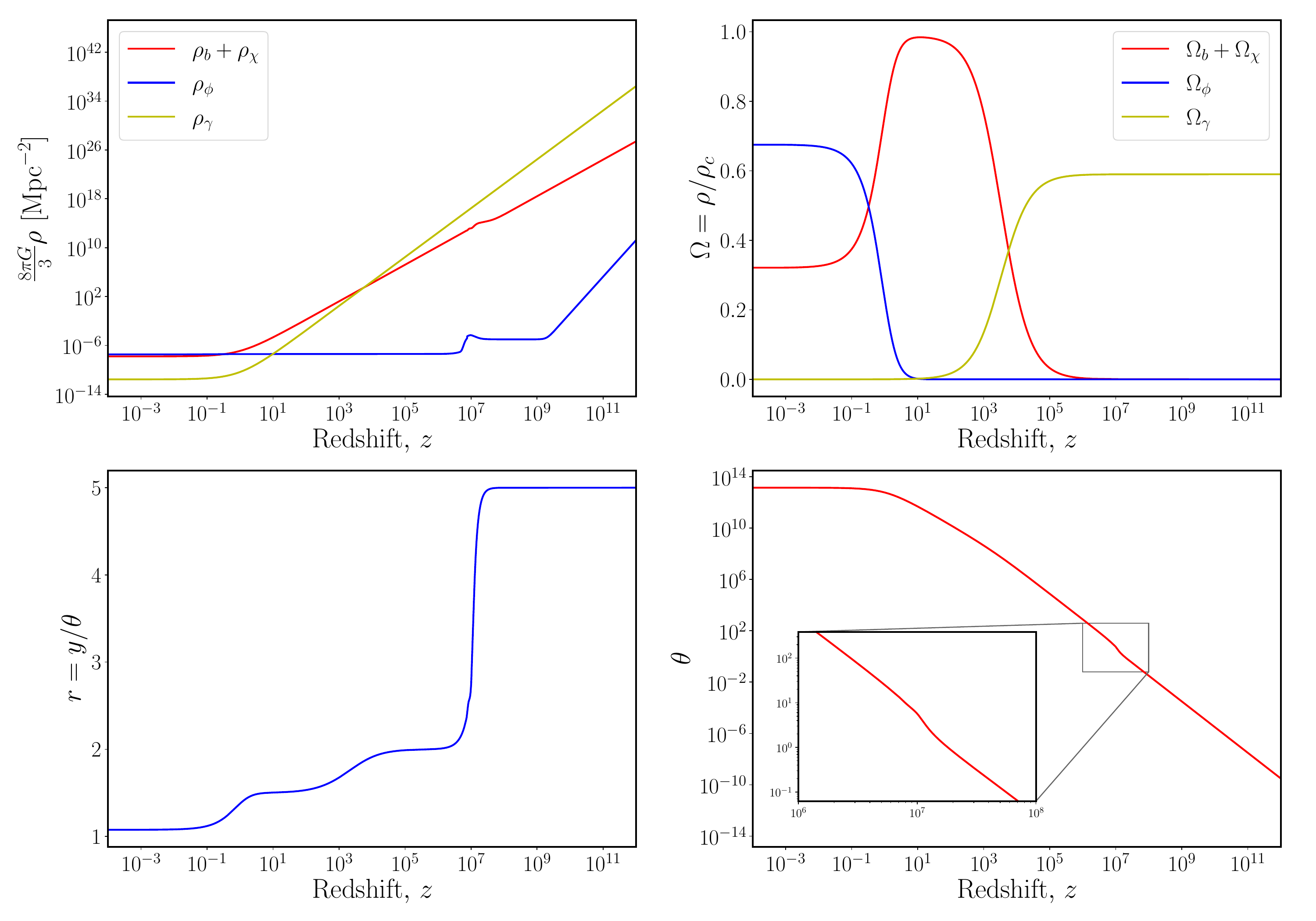}
\caption{Top panels: plot of the energy densities of matter, DE and photons (left) and the fractional energy densities (right) versus the redshift. 
Bottom panels: plot of the ratio $y/\theta$ versus $z$ (left). A sudden drop in the ratio is seen at $z\sim 10^{7}$ when fast oscillations start. Plot of $\theta$ versus $z$ (right). A sudden rise in $\theta$ is visible at around the same redshift as seen in the inset. For all panels, we take $\lambda=10^5$ Mpc$^{-2}$. }
\label{fig1}
\end{centering}
\end{figure}

The key for detecting the onset of oscillations is to track the evolution of the ratio $r=y/\theta$. We show in the bottom left panel of Fig.~\ref{fig1} the evolution of $r$ versus the redshift $z$. Oscillations begin at around $z=10^{7}$ which is where we see a sudden drop in the ratio $r$. To see the reason for this drop, analytically, we rewrite the background equations, Eqs.~(\ref{theta0}) and~(\ref{y0}) in terms of a derivative with respect to $N\equiv \ln a$, so that
\begin{align}
    \frac{\text{d}\theta}{\text{d}N}&=-3\sin\theta+y\,, \\
    \frac{\text{d}y}{\text{d}N}&=\beta \,y\,,
\end{align}
where $\beta=(3/2)(1+w_T)$ is a slowly-varying function of time. Here we are using the case with no interaction just for illustrative purposes. The rate of change of $r$ is then
\begin{equation}
    \frac{\text{d}r}{\text{dN}}=\frac{1}{\theta}\frac{\text{d}y}{\text{d}N}-\frac{y}{\theta^2}\frac{\text{d}\theta}{\text{d}N}\,,
\end{equation}
which can be written in the final form as
\begin{equation}
\frac{\text{d}r}{\text{dN}}=r\left(\beta+3\frac{\sin\theta}{\theta}-r\right)\,.    
\end{equation}
One attractor solution is
\begin{equation}
    r_0=\beta+3\frac{\sin\theta}{\theta}\,.
\end{equation}
In the limit $\theta\to 0$, $r_0\to \beta+3$ and for large $\theta$, $r_0\to \beta$. This explains the drop in $r$ when $\theta$ evolves from small to large values. During radiation domination, $w_T=1/3$, so $\beta=2$. Therefore, $y/\theta$ changes from $r_0=5$ to $2$ which is exactly what is seen in the bottom left panel of Fig.~\ref{fig1}. The bottom right panel shows the evolution of $\theta$ which exhibits a sudden rise in its value at the same value at which $r$ drops. Notice that in the upper left panel, the evolution of the DE energy density $\rho_\phi$ shows a short spike near $z=10^7$ where DM scalar field oscillations begin. This effect is communicated from DM to DE via the interaction term between the two. The upper right panel shows the evolution of the fractional density of all the species and they all follow the standard evolution with radiation domination at early times and DE domination at late times.

Define the instantaneous ratio at a given point $n$ as
\begin{equation}
    r_n=\left|\frac{y(a_n)}{\theta(a_n)}\right|\,,
\end{equation}
and the running maximum $R_n$ which is the largest ratio encountered up to and including the current point as
\begin{equation}
    R_n=\underset{1\leq i\leq n}{\text{max}}\left|\frac{y(a_i)}{\theta(a_i)}\right|\,.
\end{equation}
Once $\theta$ is sufficiently different from zero to avoid an ill-conditioned division, i.e., $|\theta|\geq \theta_{\rm floor}$, the quantity $R_n$ is updated at every accepted background point and its purpose is to estimate the value of the $y/\theta$ plateau and monitor changes in its evolution. A candidate drop in the value of $r_n$ is identified when two conditions are met:
\begin{align}
  r_n &< r_{n-1},\\
  r_n &\leq f_R R_{n}\,,
\end{align}
where we chose $f_R=0.70$ which means that the ratio must fall by at least $30\%$ below its running maximum. The first condition requires the ratio be decreasing at the current background point. The second requires that the decrease be sufficiently large relative to the previously learned plateau. The first accepted background point satisfying both conditions is stored and the value of $\theta$ at this point is $\theta_{\rm drop}$. 

This ratio criterion is not sufficient by itself since a local increase in $\theta$ can create a decrease in $r$ which the algorithm may classify as a drop due to the onset of rapid oscillations. Therefore, an independent test of the sign changes in the equation of state is implemented. 

We have seen that the onset of fast oscillations brings about a rapidly varying equation of state with alternating sign. The code tracks the value of $w_\chi$ and at each accepted background point, the sign of $w_{\chi}$ is classified according to
\begin{equation}
s_n=
\begin{cases}
+1, & w_{\chi}\geq w_{\min},\\
-1, & w_{\chi}\leq-w_{\min},\\
0,  & |w_{\chi}|<w_{\min}.
\end{cases}
\end{equation}
The finite amplitude $w_{\min}$ defines a narrow corridor around zero. Values inside this corridor are ignored so that numerical noise or a single point located close to a zero crossing is not counted as a physical change of sign. Whenever a new nonzero sign differs from the previous nonzero sign, the sign-change counter is increased. Samples with $s_n=0$ neither increase the counter nor erase the most recently resolved sign. The oscillatory behavior is regarded as confirmed only after all of the following have occurred:
\begin{enumerate}
    \item at least one positive value of $w_\chi$,
    \item at least one negative value of $w_\chi$,
    \item the number of sign changes must be greater than or equal some number which is set to 2 by default.
\end{enumerate}
Requiring more than one sign change provides additional evidence that the phase is genuinely running through repeated oscillations rather than undergoing a single crossing.

Once the $w_\chi$ test confirms the interpretation of the stored ratio drop, the switching phase is set to
\begin{equation}
\theta_{*}=\max\left(\theta_{\rm drop}+\Delta\theta_{\rm pad},\theta_{\min}\right)\,,
\end{equation}
where $\theta_{\rm min}$ is the minimum $\theta$ value that must be reached before the switch trigger can be accepted and the padding $\Delta\theta_{\rm pad}$ allows the exact solution to continue for a short additional phase interval after the detected onset. This ensures that the code does not begin averaging at the very first indication of the transition.

\begin{figure}[t]
\begin{centering}
\includegraphics[width=1.0\linewidth]{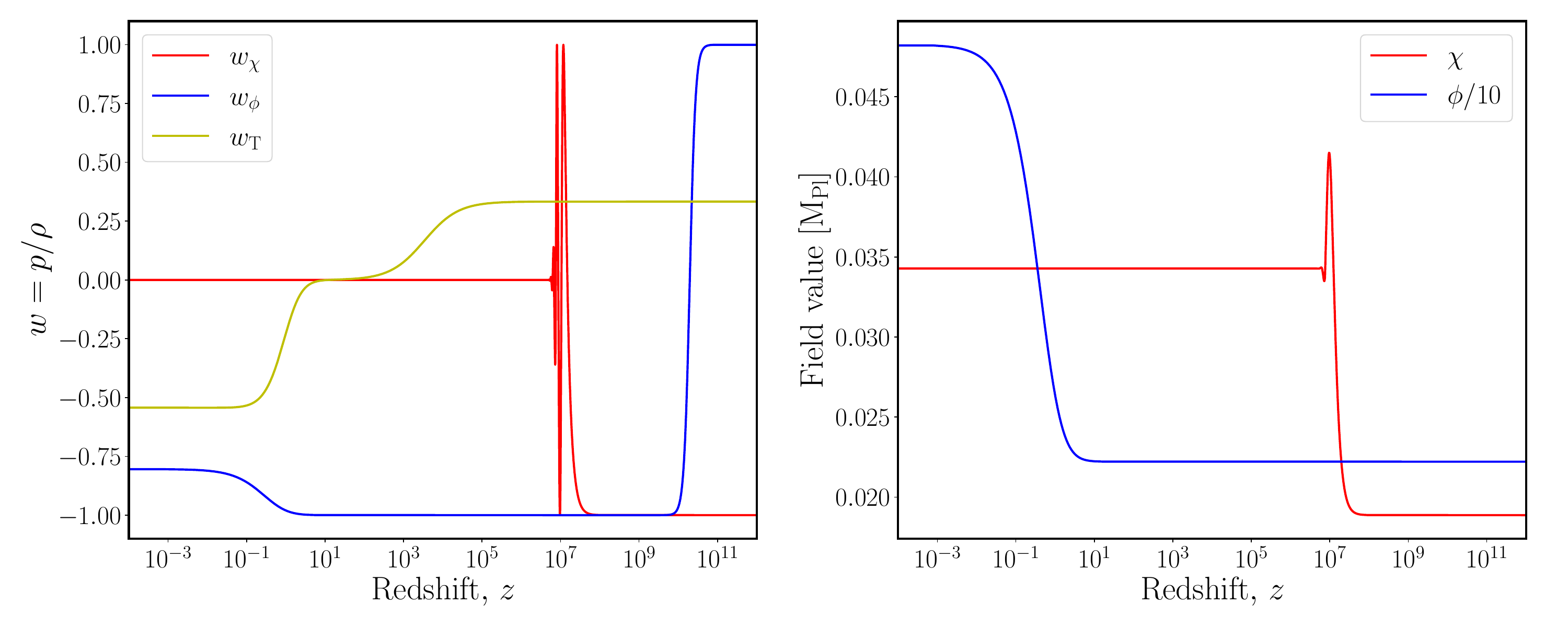} \\
\includegraphics[width=1.0\linewidth]{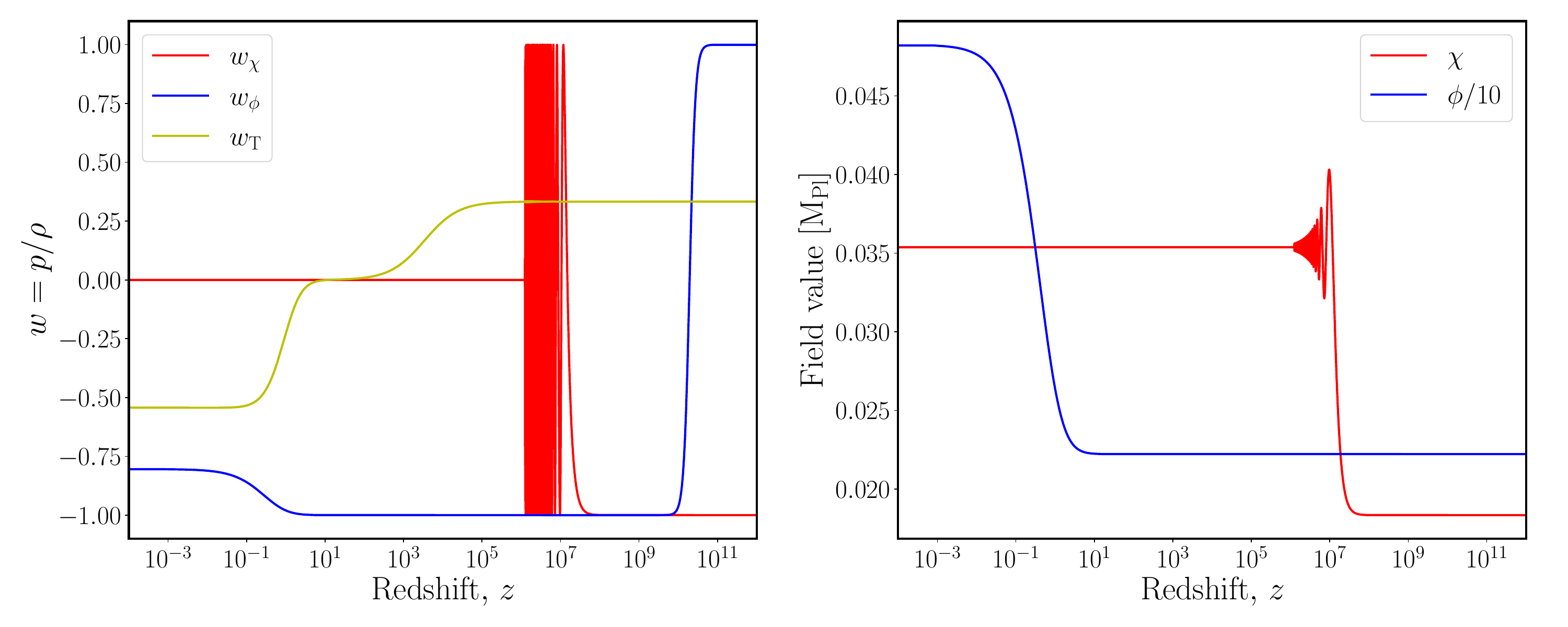}
\caption{Plots of the DM and DE equations of state and the total equation of state (left) and the evolution of the DM field $\chi$ (right) as a function of $z$. The top panels are generated using the `automatic' mode while the bottom panels use the `manual' mode. For all panels, we take $\lambda=10^5$ Mpc$^{-2}$. }
\label{fig2}
\end{centering}
\end{figure}

In Fig.~\ref{fig2} we show the effect of the `automatic' (upper panel) and `manual' (lower panel) modes on the DM equation of state. It is clear that the `automatic' mode precisely determines the redshift at which oscillations begin and trigger the averaging procedure, whereas in the `manual' mode, the field undergoes rapid oscillations for a longer period of time before the averaging is triggered. One can see the same effects on the evolution of the DM field $\chi$ in the right panels of Fig.~\ref{fig2}.

We show in table~\ref{tab1} the controls implemented in \code{CLASS} for the automatic detection of the oscillatory phase of the scalar field. The key \code{osc\_threshold\_mode} can be set to either `automatic' or `manual'. In the `automatic' mode, the code detects the transition based on the mechanism described here while in the `manual' mode, the code utilizes the user-specified threshold which is the classical treatment.

\begin{table}[ht]
\centering
\begin{tabular}{>{\ttfamily}l l p{7.0cm}}
\toprule
\normalfont Parameter & Typical value & Role \\
\midrule
osc\_threshold\_mode & automatic & Selects the automatic detector. \\
osc\_transition\_width & 5 & Width $w$ of the $\tanh$ interpolation. \\
osc\_auto\_ratio\_theta\_floor & $10^{-12}$ & Phase floor used while learning $R$. \\
osc\_auto\_min\_theta & 6 & Minimum phase before the trigger can be accepted. \\
osc\_auto\_ratio\_drop & 0.70 & Fraction $f_R$ defining the required drop. \\
osc\_auto\_w\_amplitude & 0.10 & Minimum resolved $|w_{\chi}|$. \\
osc\_auto\_min\_sign\_changes & 2 & Required number of nonzero sign changes. \\
osc\_auto\_padding & 5 & Padding added to $\theta_{\rm drop}$. \\
\bottomrule
\end{tabular}
\caption{Automatic-mode numerical controls.}
\label{tab1}
\end{table}

In the `automatic' mode, an \code{.ini} file block of \code{CLASS} looks like:
\begin{lstlisting}[style=classcode,language={}]
osc_threshold_mode = automatic
osc_transition_width = 5.0
osc_auto_ratio_theta_floor = 1e-12
osc_auto_min_theta = 6.0
osc_auto_ratio_drop = 0.70
osc_auto_w_amplitude = 0.10
osc_auto_min_sign_changes = 2
osc_auto_padding = 5.0
\end{lstlisting}
whereas the manual mode looks like:
\begin{lstlisting}[style=classcode,language={}]
osc_threshold_mode = manual
osc_threshold = 100.0
\end{lstlisting}

The modified \code{CLASS} code used in this analysis is made available on the \href{https://github.com/aabouibrahim/class_field_theoretic}{Github} page. To locate the changes made to \code{CLASS}, users can search for the word \code{NEW} in the \code{input.c}, \code{background.c} and \code{perturbations.c} modules and the corresponding header files.

\section{Effect of DM-DE interaction on perturbations}\label{sec:perturb}

\begin{figure}[t]
\begin{centering}
\includegraphics[width=0.49\linewidth]{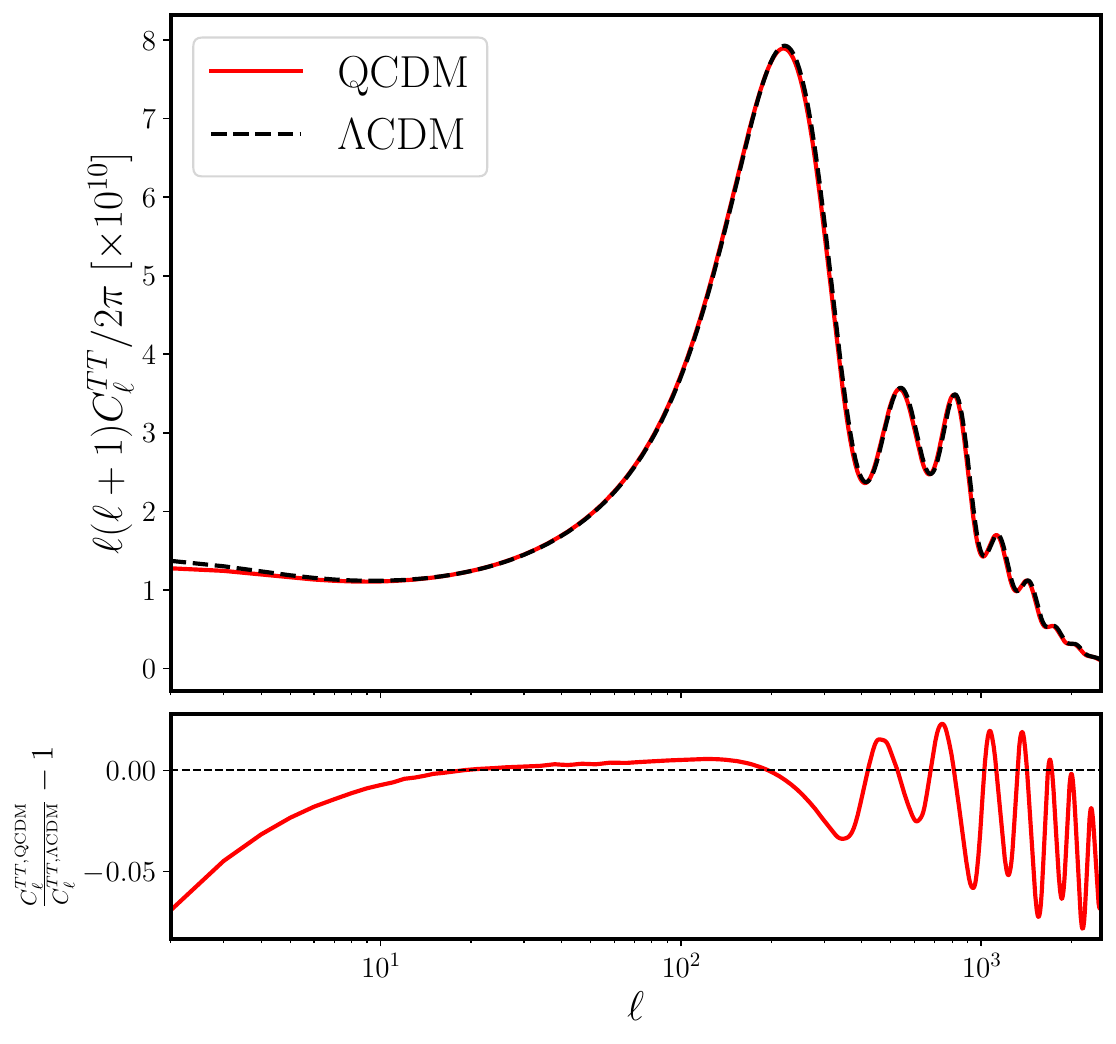}
\includegraphics[width=0.49\linewidth]{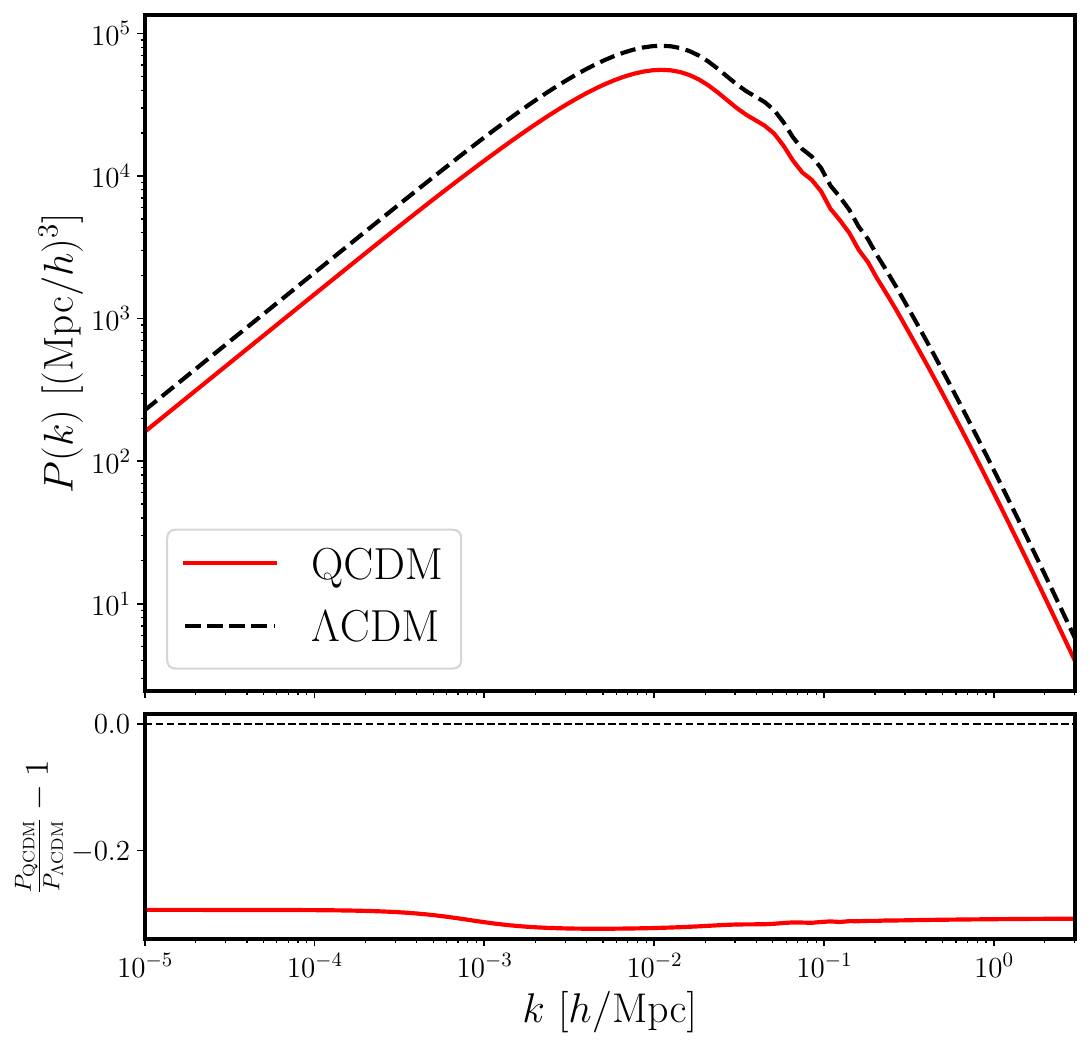}
\caption{Left panel: The temperature-temperature correlations in the lensed CMB power spectrum showing QCDM (solid line) and $\Lambda$CDM (dashed). Right panel: the matter power spectrum for QCDM and $\Lambda$CDM. For this benchmark, $\lambda=10^5$ Mpc$^{-2}$.}
\label{fig3}
\end{centering}
\end{figure}

In this section, we study the effect of DM-DE interaction on observables tied to linear perturbations, namely, the CMB and matter power spectra. The lensed temperature-temperature power spectrum of the CMB is shown in the left panel of Fig.~\ref{fig3} for QCDM (solid curve) and $\Lambda$CDM (dashed curve). The bottom panel shows the difference in prediction between the two models. It is clear that the overall CMB power spectrum of QCDM matches closely that of $\Lambda$CDM on almost all scales, where the acoustic peak amplitudes are largely unaltered which shows that DM-DE interaction does not alter the physics of recombination. However, examining the lower panel for minute changes, one can see that QCDM features a power suppression at the lowest multipoles, an oscillatory pattern for higher multipoles and a percent-level change in the first acoustic peak. The interesting modulating fractional change seen at high $\ell$ suggests that DM-DE interaction modifies the gravitational driving and damping of the photon-baryon system.  

\begin{figure}[H]
\begin{centering}
\includegraphics[width=0.49\linewidth]{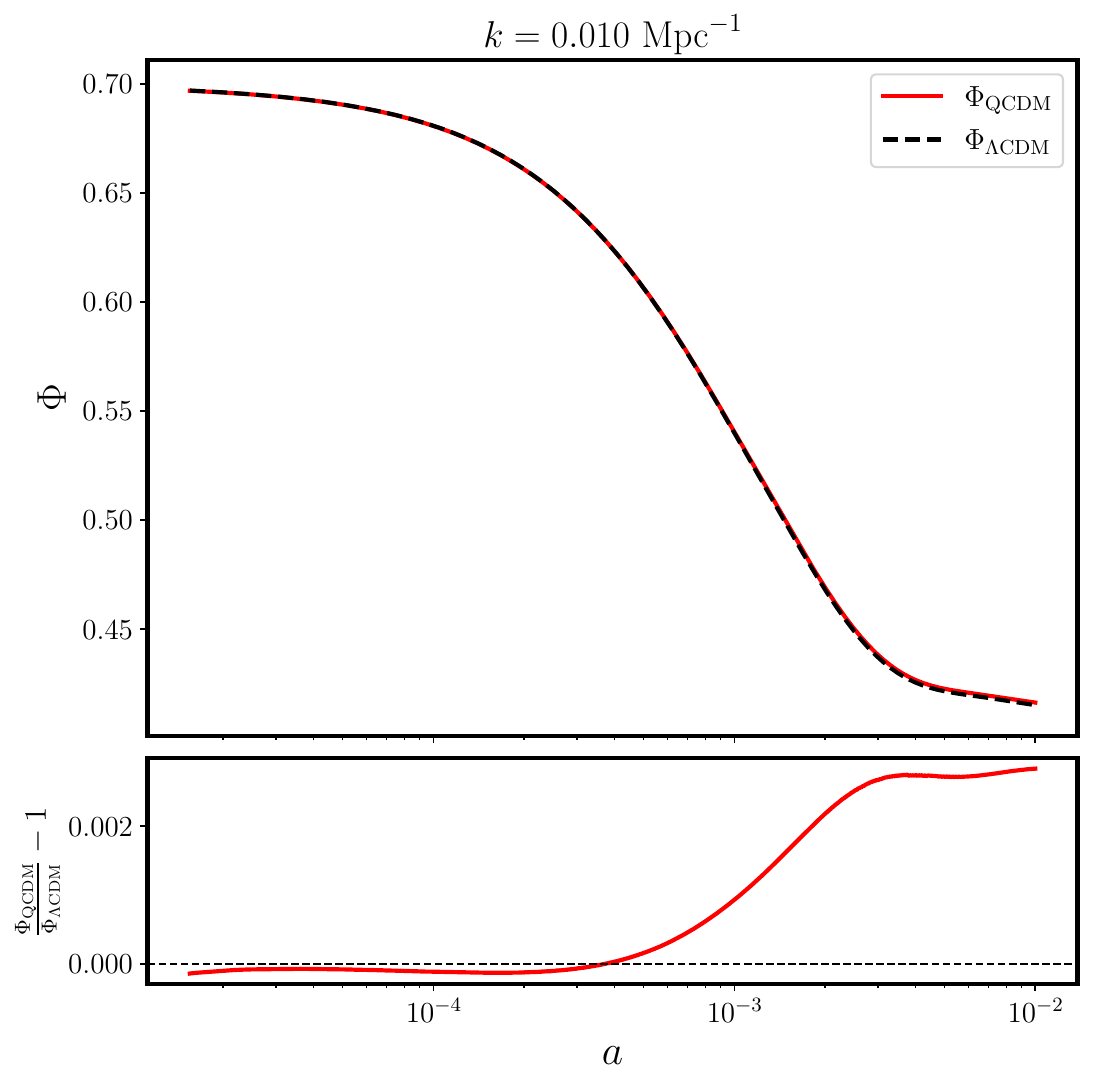}
\includegraphics[width=0.49\linewidth]{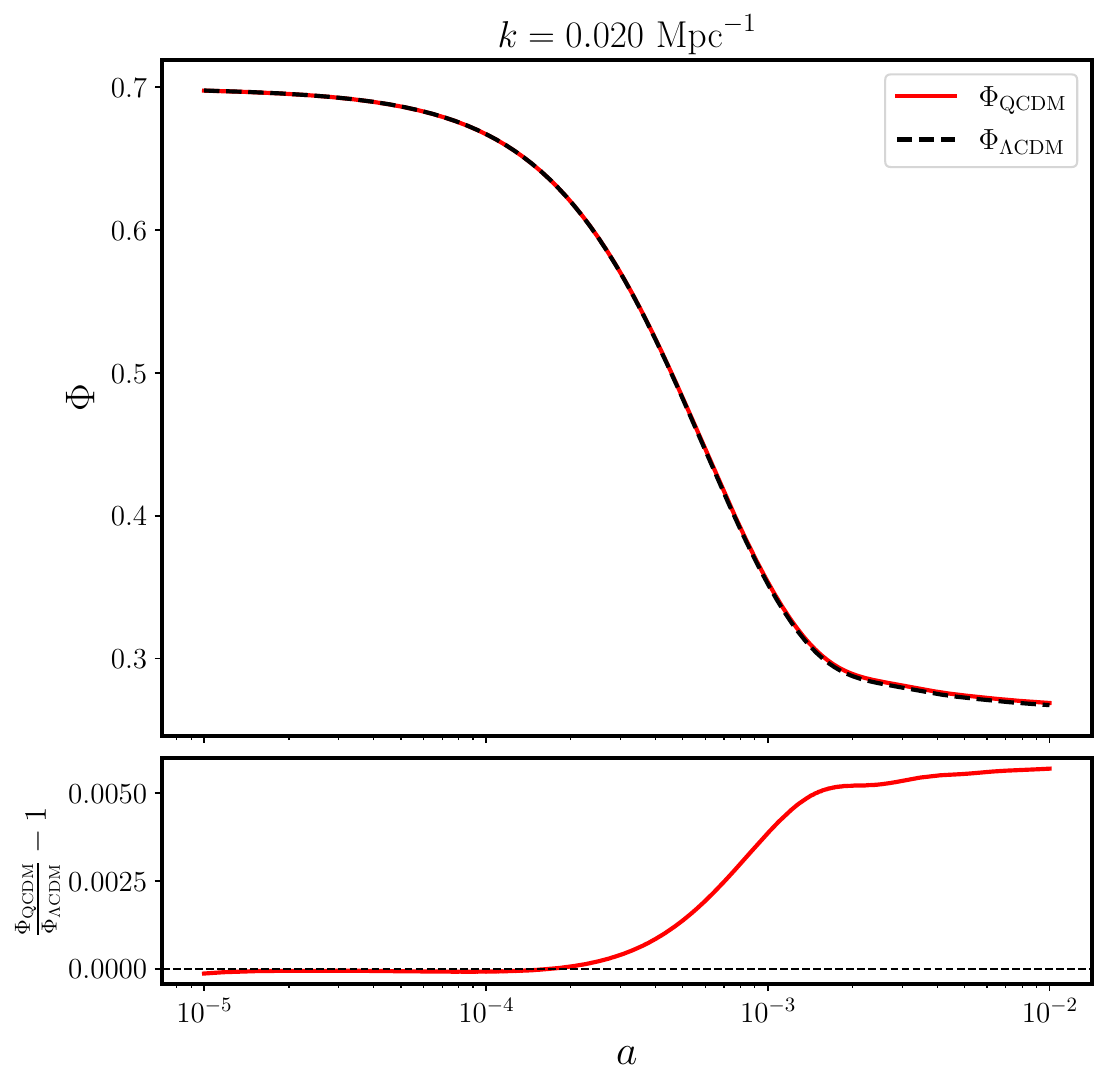} \\
\includegraphics[width=0.49\linewidth]{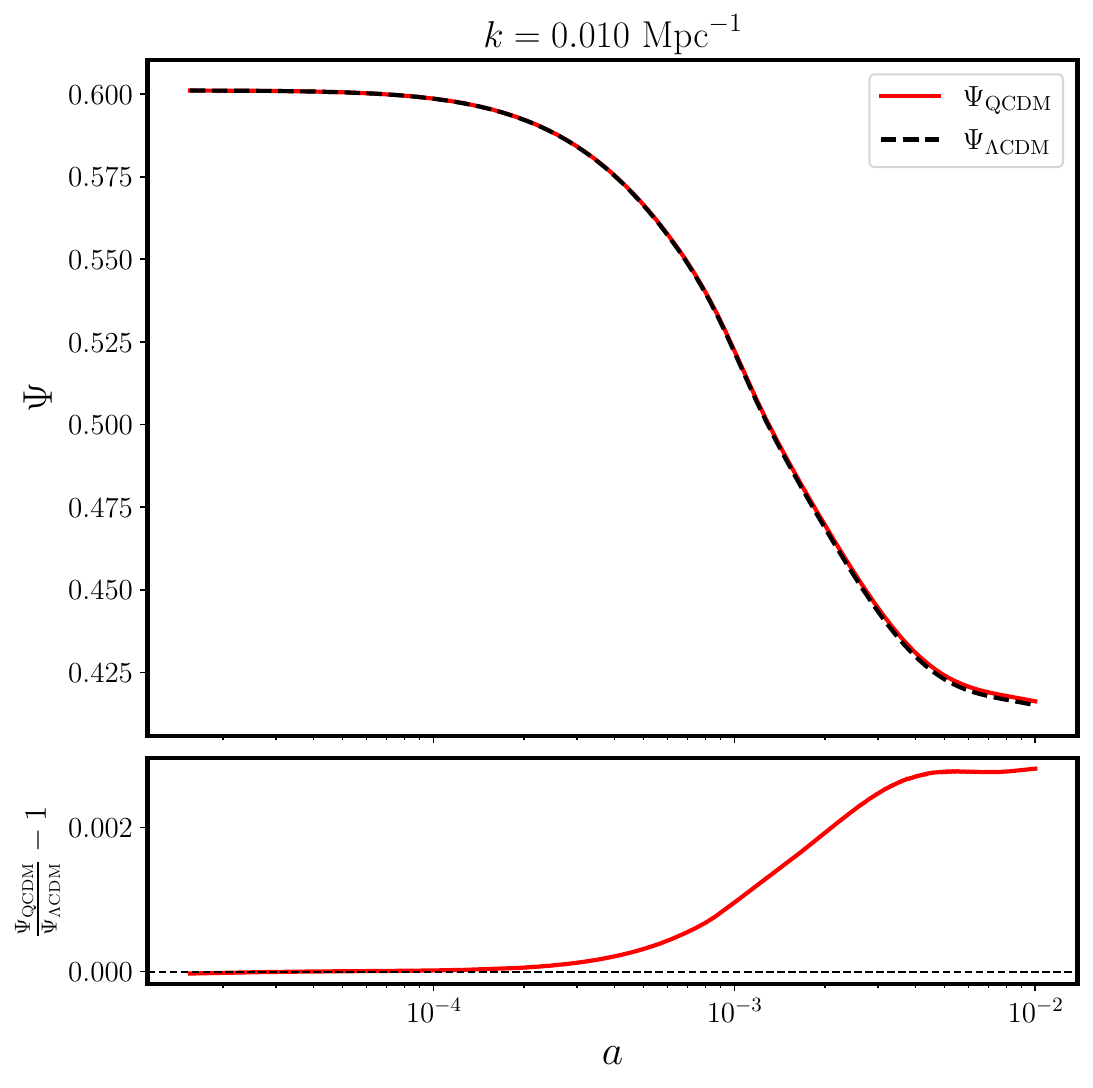}
\includegraphics[width=0.49\linewidth]{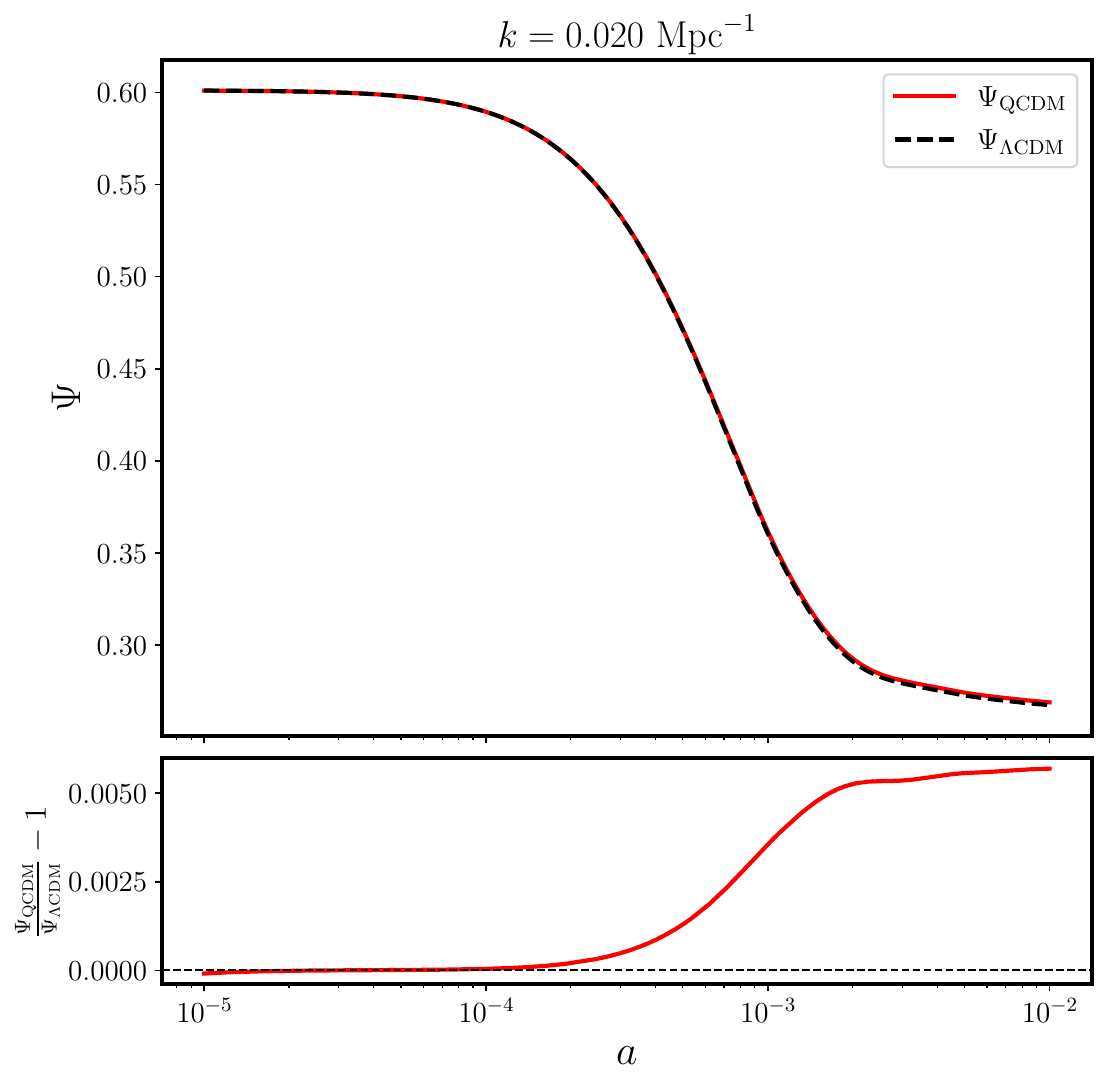}
\caption{Plots of the Newtonian potentials, $\Phi$ (upper panel) and $\Psi$ (lower panel) for two wavenumbers as a function of the scale factor. The fractional difference between QCDM and $\Lambda$CDM is also shown in each panel. For this benchmark, $\lambda=10^5$ Mpc$^{-2}$.  }
\label{fig4}
\end{centering}
\end{figure}

To verify the above interpretation, we plot in Fig.~\ref{fig4} the Newtonian potentials $\Phi$ and $\Psi$ as a function of the scale factor for two values of the wavenumber. These modes are the ones relevant to the first few acoustic peaks and they tell a consistent story. QCDM produces an enhancement in the gravitational potential for increasing values of $a$ and this enhancement is more pronounced for higher values of $k$. Therefore, the interaction changes the evolution of the metric potentials ever so slightly around recombination which influences the gravitational driving of the photon-baryon oscillator. This naturally accounts for the small oscillatory features seen in the CMB power spectrum. 

\begin{figure}[H]
\begin{centering}
\includegraphics[width=0.49\linewidth]{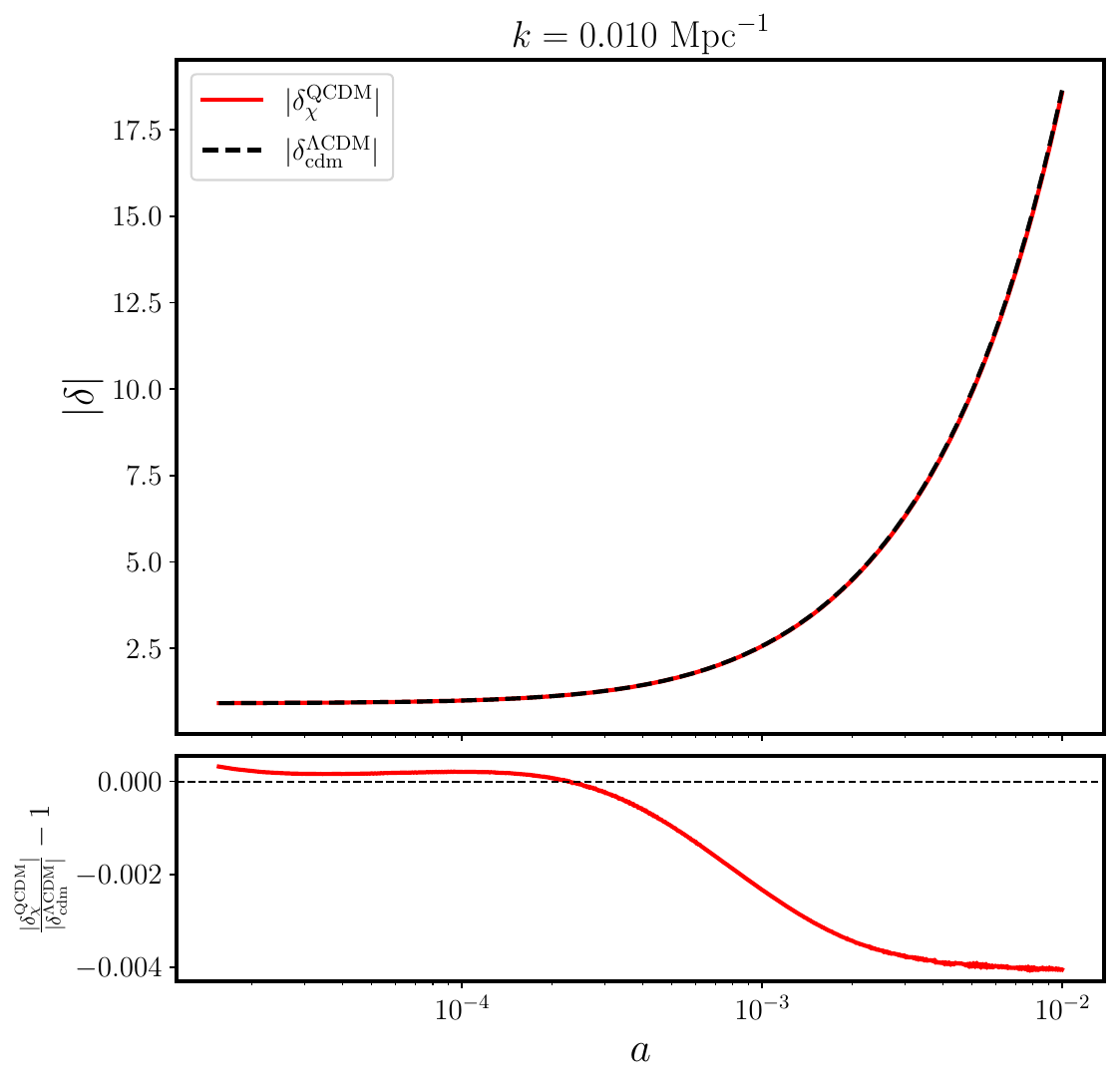}
\includegraphics[width=0.49\linewidth]{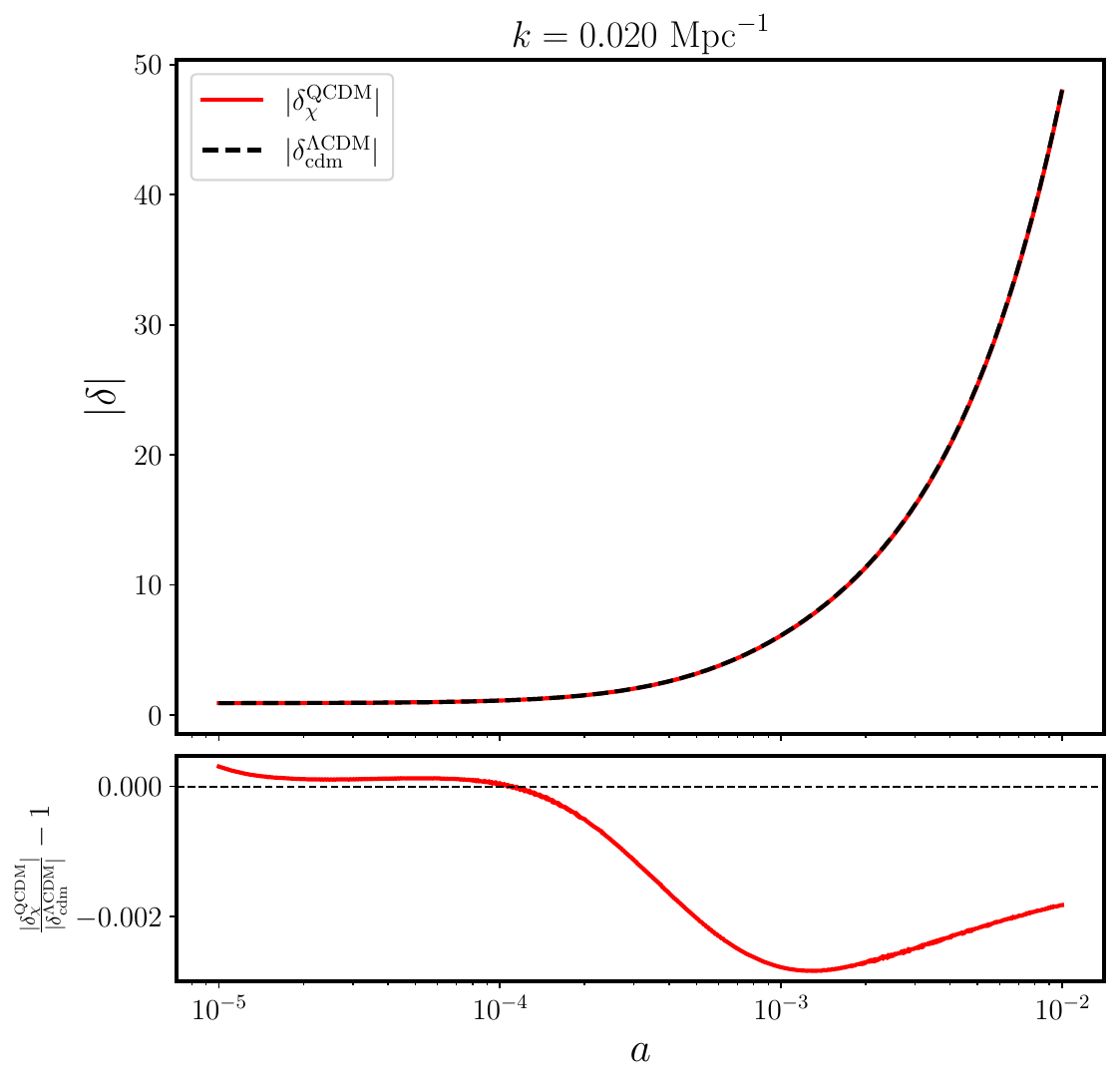}
\caption{Plots of the density contrasts of DM in the QCDM model (solid line) and $\Lambda$CDM model (dashed line) for two values of the wavenumber. For this benchmark, $\lambda=10^5$ Mpc$^{-2}$.  }
\label{fig5}
\end{centering}
\end{figure}

Now, we turn our attention to the matter power spectrum. To understand this, we examine the growth of the scalar field DM density contrast $\delta_\chi$ in QCDM and compare it to $\delta_{\rm cdm}$ in $\Lambda$CDM, as shown in Fig.~\ref{fig5}. It is clear that scalar field DM follows very closely CDM at early times and only begins to show deviations below the percent level just after $a\sim 10^{-4}$. This means that any suppression of the matter power spectrum seen in QCDM is cumulative and does not happen instantaneously near recombination. This leads us to the most prominent signature of DM-DE interaction reflected in the suppression of the matter power spectrum at late times over a range of modes as seen in the right panel of Fig.~\ref{fig3}. QCDM predicts a suppression relative to $\Lambda$CDM by roughly $30\%-35\%$, with the largest dip around $k\sim10^{-3}-10^{-2}\,h\,{\rm Mpc}^{-1}$, followed by a mild increase at higher wavenumbers. This smooth suppression over all modes is indicative of a cumulative suppression in growth rather than a sudden cutoff in $|\delta_\chi|$.

There is an interesting distinction between the CMB and large-scale structure responses. The CMB is mostly established at \(z\simeq1100\), when the QCDM perturbations and potentials still differ from \(\Lambda\)CDM by only a fraction of a percent. Consequently, the primary TT spectrum receives only modest changes. Matter perturbations, on the other hand, continue to evolve for the entire interval from recombination to the present. Even a relatively small modification of the growth rate,
\begin{equation}
f(k,a)\equiv\frac{\text{d}\ln\delta_m}{\text{d}\ln a},    
\end{equation}
can therefore integrate into a large difference in the final amplitude:
\begin{equation}
\frac{\delta_{\rm QCDM}(a_0)}{\delta_{\Lambda{\rm CDM}}(a_0)}=\exp\left[\int\left(f_{\rm QCDM}-f_{\Lambda{\rm CDM}}\right)\text{d}\ln a\right].    
\end{equation}
Since $P\propto\delta^2$, a 30\% reduction in $P$ corresponds only to an amplitude ratio of approximately
$\delta_{\rm QCDM}/\delta_{\Lambda\text{CDM}}\simeq 0.84$, i.e. about a 16\% reduction in the present-day density-contrast amplitude. Such a difference can plausibly accumulate gradually over many e-folds of late-time evolution without requiring a comparably large departure around recombination.

\section{Updated constraints on cosmological parameters}\label{sec:constraints}

The Boltzmann solver \code{CLASS} is interfaced with \code{Cobaya}~\cite{Torrado:2020dgo}, a code for sampling and statistical modeling, to perform a Markov Chain Monte Carlo (MCMC) analysis of the model parameter space which allows us to extract constraints on the cosmological parameters through Bayesian inference. \code{Cobaya} utilizes an adaptive, speed-hierarchy-aware MCMC sampler (adapted from \code{CosmoMC})~\cite{Lewis:2002ah,Lewis:2013hha} using the fast-dragging procedure described in ref.~\cite{Neal:2005uqf}. We monitor the convergence of the chains using the Gelman-Rubin~\cite{Gelman:1992zz} criterion $R-1<0.02$. After convergence, the chains are analyzed with \code{GetDist}~\cite{Lewis:2019xzd}\footnote{\url{https://github.com/cmbant/getdist}}, a package allowing for the extraction of numerical results, including 1D posteriors and 2D marginalized probability contours. In this section we update the constraints on the QCDM model parameters using the most recent public data. The data sets are organized as follows: 

\begin{enumerate}

\item \textbf{DESI DR2 BAO}: We use baryon acoustic oscillation (BAO) measurements from the second data release of the Dark Energy Spectroscopic Instrument (DESI DR2), based on galaxies, quasars, and the Ly$\alpha$ forest over the approximate redshift range $0.1<z<4.2$. These measurements constrain the late-time expansion history through the transverse and radial BAO distance scales~\cite{DESI:2025zgx,DESI:2025zpo}. This data set is referred to as \textbf{DESI}.

\item \textbf{Pantheon+ supernovae}: We use the Pantheon+ compilation of Type Ia supernovae, consisting of 1701 light curves from 1550 SNe Ia spanning $0.001<z<2.26$. The sample constrains the relative luminosity-distance--redshift relation and hence the late-time expansion history. We use the Pantheon+ likelihood without the SH0ES Cepheid calibration~\cite{Brout:2022vxf}. This data set is referred to as \textbf{PP}.

\item \textbf{Pantheon+SH0ES supernovae}: We also consider the Pantheon+SH0ES likelihood, which combines the Pantheon+ Type Ia supernova sample with the SH0ES Cepheid calibration of supernova host galaxies. In contrast to Pantheon+ alone, the inclusion of the distance-ladder calibration provides an absolute calibration of the supernova luminosities and therefore directly constrains the present-day Hubble constant $H_0$~\cite{Brout:2022vxf}. This data set is referred to as \textbf{PPS}.

\item The CMB data used is under the name \textbf{CMB} and is comprised of:
\begin{enumerate}
\item \textbf{Planck 2018 primary CMB}: We use the Planck 2018 temperature and polarization power spectra, including the low-$\ell$ TT likelihood and the Planck TT, TE, and EE spectra at higher multipoles. For consistency with the ACT combination, we restrict the Planck spectra to $\ell_{\rm max}^{TT}=1000$ and $\ell_{\rm max}^{TE,EE}=600$~\cite{Aghanim:2019ame,AtacamaCosmologyTelescope:2025nti}.

\item \textbf{ACT DR6 primary CMB}: We include the ACT DR6 high-resolution TT, TE, and EE power spectra over the multipole range $600\leq\ell\leq8500$. These measurements complement Planck by providing precise information on the small-scale CMB and the acoustic damping tail~\cite{AtacamaCosmologyTelescope:2025nti,AtacamaCosmologyTelescope:2025blo}.

\item \textbf{ACT DR6 + Planck PR4 CMB lensing}: We use the ACT DR6 lensing likelihood in its \texttt{actplanck\_baseline} configuration, which combines the ACT DR6 and Planck PR4 CMB lensing reconstructions while accounting for their correlations. The lensing spectrum provides an important probe of the integrated matter distribution and the growth of cosmic structure~\cite{Sailer:2024jrx,ACT:2023kun,ACT:2023dou,Carron:2022eyg}.

\item \textbf{SPT-3G D1 primary CMB}: We include the SPT-3G D1 temperature and polarization measurements from the 2019--2020 observations of the SPT-3G Main field. We use the ``lite'' likelihood with $400<\ell<3000$ for TT and $400<\ell<4000$ for TE and EE, providing an independent high-resolution measurement of the small-scale CMB~\cite{SPT-3G:2025bzu}.

\item \textbf{SPT-3G CMB lensing}: We include the SPT-3G CMB lensing reconstruction obtained from the 2019--2020 polarization data using the MUSE analysis. In our likelihood, we use only the reconstructed lensing-potential power spectrum, $C_L^{\phi\phi}$, which provides an independent constraint on the projected matter distribution and the growth of structure~\cite{SPT-3G:2024atg,Millea:2021had}.
\end{enumerate}

\end{enumerate}

We impose flat priors on the model parameters (shown in table~\ref{tab-prior}), except for $\tau_{\rm reio}$ where a Gaussian prior is implemented with a mean value of 0.051 and standard deviation of 0.006. 

\begin{table}[ht]
\centering
{\tabulinesep=1.2mm
\resizebox{0.3\textwidth}{!}{\begin{tabu}{cc}
\hline\hline
\textbf{Parameter} & \textbf{Prior} \\
\hline
$\log(10^{10}A_{s})$ & $[1.61, 3.91]$ \\
$n_{s}$ & $[0.8, 1.2]$ \\
$H_0$ & $[40, 100]$ \\
$\omega_b$ & $[0.005, 0.1]$ \\
$\omega_\chi$ & $[0.001, 0.99]$ \\
{$F~[M_{\rm Pl}]$} & $[0.1,1.0]$ \\
{$\phi_{\mathrm{ini}}$} & $[0.01,0.99]$ \\
{$\log(\lambda/\text{Mpc}^{-2})$} & $[-2.0,9.0]$ \\
\hline\hline
\end{tabu}}}
\caption{The ranges of the flat priors adopted for the cosmological parameters in the QCDM model. The input DM mass is taken to be $m_\chi=10^{-20}$ eV. }
\label{tab-prior}
\end{table}

The parameter $\mu^4$ that enters in the DE quintessence potential becomes a derived parameter. In fact, we use $\log\mu^4$ in \code{CLASS} as a shooting parameter which gets updated starting from a guessed value and the algorithm predicts the correct value that satisfies the closure condition $\sum\Omega_i=1$, where the sum is over all species.

We use \code{MCEvidence}\footnote{\url{https://github.com/yabebalFantaye/MCEvidence}}~\cite{Heavens:2017hkr,Heavens:2017afc} to calculate the Bayes factors $\ln\mathcal{B}_{ij}$ normalized to the baseline $\Lambda$CDM which allows us to compare the performance of QCDM versus $\Lambda$CDM in fitting the data. A negative value means that data favors $\Lambda$CDM while a positive value means QCDM is favored. The classification uses the revised Jeffreys scale by Kass and Raftery as given in  refs.~\cite{Kass:1995loi,Trotta:2008qt}: (i) for $0 \leq | \ln \mathcal{B}_{ij}|  < 1$, the model has an inconclusive evidence, (ii) for $1 \leq | \ln \mathcal{B}_{ij}|  < 2.5$, the model has a weak evidence, (iii) for $2.5 \leq | \ln \mathcal{B}_{ij}|  < 5$, the model has a moderate evidence, and, finally, (iv)  for $ | \ln \mathcal{B}_{ij} | \geq 5$, the model has a strong evidence.

Table~\ref{tab3} summarizes the constraints on the QCDM parameters for the three data combinations. The inclusion of DESI and the additional low-redshift data significantly tightens the parameter constraints and shifts the preferred value of the Hubble constant upward, from $H_0=66.5^{+1.3}_{-0.70}\,\mathrm{km\,s^{-1}\,Mpc^{-1}}$ for CMB alone to $H_0=68.37^{+0.30}_{-0.26}\,\mathrm{km\,s^{-1}\,Mpc^{-1}}$ for CMB+DESI+PPS. This is accompanied by a decrease in both $\Omega_{\rm m}$ and $S_8$, with the latter reaching $S_8=0.8119\pm0.0079$ for the full data combination. The additional datasets also strengthen the bounds on the scalar-field parameters, particularly $F$ and $\phi_{\rm ini}$, while the interaction parameter $\lambda$ remains comparatively weakly constrained.

\begin{table}[t]
\centering
{\tabulinesep=1.2mm
\resizebox{\textwidth}{!}{\begin{tabu}{ccccc}
\hline\hline
\textbf{Parameter} & \textbf{CMB} & \textbf{CMB+DESI+PP} & \textbf{CMB+DESI+PPS} \\
\hline
$\log(10^{10} A_\mathrm{s})$ & $3.051\pm 0.010$ & $3.0622\pm 0.0097$ & $3.0647\pm 0.0097$ \\
{$n_\mathrm{s}$} & $0.9710\pm 0.0033$ & $0.9757\pm 0.0029$ & $0.9766\pm 0.0029$ \\
{$H_0~\mathrm{[km/s/Mpc]}$} & $66.5^{+1.3}_{-0.70}$ & $67.74^{+0.52}_{-0.39}$ & $68.37^{+0.30}_{-0.26}$ \\
{$\tau_\mathrm{reio}$} & $0.0548\pm 0.0055 $ & $0.0599\pm 0.0053$ & $0.0610\pm 0.0054$ \\
{$\Omega_\mathrm{b} h^2$} & $0.022439\pm 0.000095$ & $0.022519\pm 0.000093$ & $0.022570\pm 0.000095$ \\
{$\Omega_\chi h^2$} & $0.12013\pm 0.00096$ & $0.11792\pm 0.00064$ & $0.11739\pm 0.00058$ \\
{$F~\mathrm{[M_{Pl}]}$} & $> 0.649$ & $> 0.683$ & $> 0.704$ \\
{$\phi_{\mathrm{ini}}~\mathrm{[M_{Pl}]}$} & $< 0.500$ & $< 0.492$ & $< 0.249$ \\
{$\log(\lambda/\mathrm{Mpc}^{-2})$} & $2.2^{+1.8}_{-3.5}$ & $2.2^{+2.0}_{-3.4}$ & $< 3.99$ \\
\hline 
$\Omega_\mathrm{m}$ & $0.3243^{+0.0082}_{-0.014} $ & $0.3077^{+0.0042}_{-0.0051}$ & $0.3010\pm 0.0033$ \\
$\Omega_\phi$ & $0.676^{+0.014}_{-0.0082}$ & $0.6922^{+0.0051}_{-0.0042}$ & $0.6989\pm 0.0033$ \\
$\log\mu^4$ & $-7.232^{+0.038}_{-0.075}$ & $-7.225^{+0.029}_{-0.049}$ & $-7.248^{+0.011}_{-0.020}$ \\
$S_8$ & $0.837^{+0.011}_{-0.0094}$ & $0.8179^{+0.0071}_{-0.0062}$ & $0.8119\pm 0.0079$ \\
\hline
$\Delta\chi^2_{\rm min}$ & $+1.253$ & $-1.307$ & $+0.657$ \\
$\ln\mathcal{B}_{ij}$ & $-0.569$ & $-0.861$ & $-2.189$ \\
\hline\hline
\end{tabu}}}
\caption{Constraints on some of the cosmological parameters of the QCDM model. The values are quoted at 68\% CL intervals for three data set combinations. The middle line separates the sampled from the derived parameters using MCMC. In the last two rows we show the values of $\Delta\chi^2_{\rm min}\equiv\chi^2_{\rm QCDM,min}-\chi^2_{\rm \Lambda CDM,min}$ and the Bayes factor $\ln\mathcal{B}_{ij}$. }
\label{tab3}
\end{table}

The goodness-of-fit comparison shows no substantial improvement of QCDM over $\Lambda$CDM. Although CMB+DESI+PP yields a modest improvement, with $\Delta\chi^2_{\rm min}=-1.307$, the CMB-only and CMB+DESI+PPS combinations give slightly larger minimum $\chi^2$ values than $\Lambda$CDM. Moreover, the negative Bayes factors for all three data combinations indicate an overall Bayesian preference for $\Lambda$CDM, with the preference becoming stronger for the full CMB+DESI+PPS dataset, for which $\ln\mathcal{B}_{ij}=-2.189$. Thus, while QCDM provides a viable fit to the data and produces modest shifts in several cosmological parameters, the additional model freedom is not sufficiently favored by the data to overcome the Bayesian complexity penalty.

\begin{figure}[H]
\begin{centering}
\includegraphics[width=\linewidth]{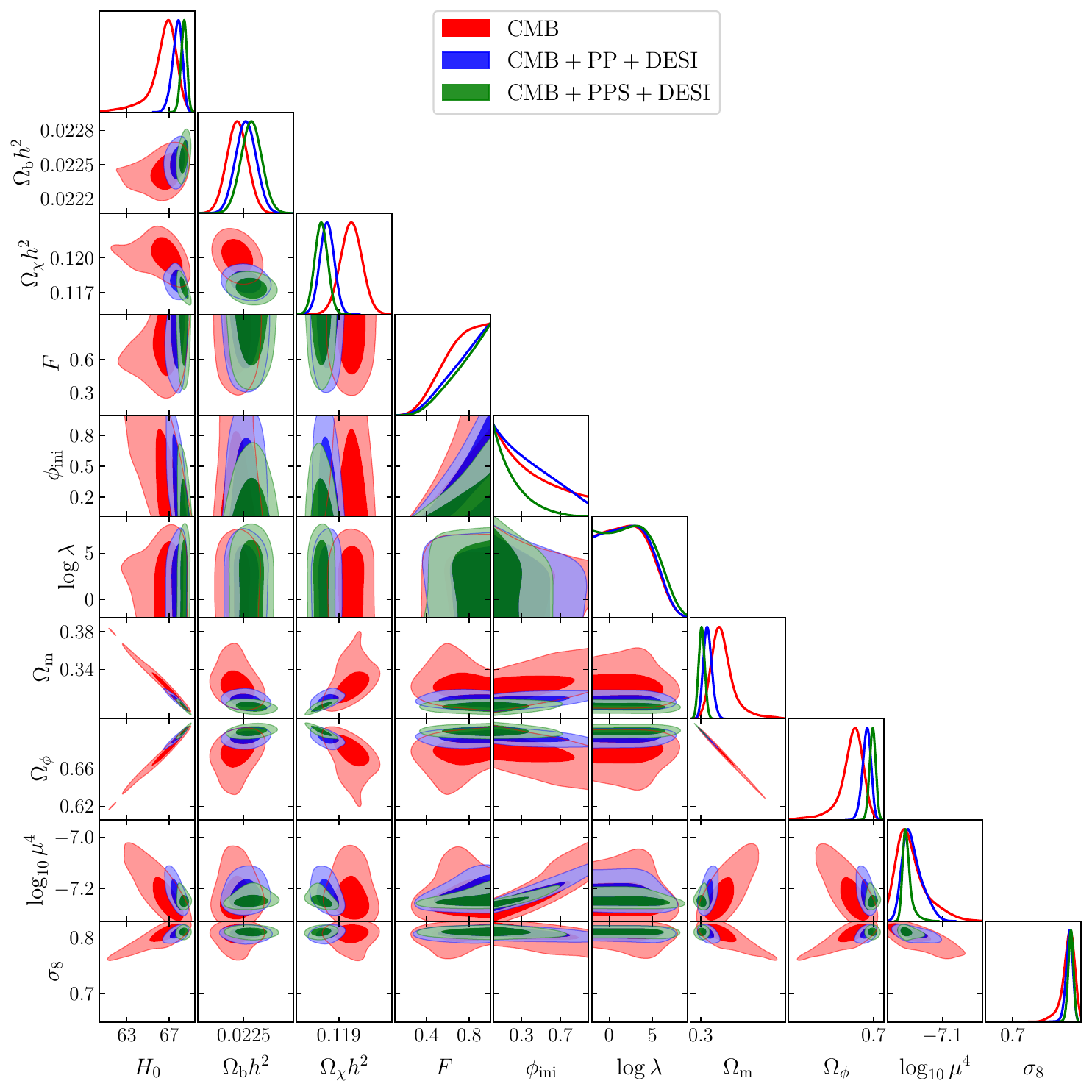}
\caption{Triangle plot showing the 2D joint and 1D marginalized posteriors of a number cosmological parameters of QCDM for the following data set combinations: CMB (red contours), CMB+PP+DESI (blue contours) and CMB+PPS+DESI (green contours). }
\label{fig6}
\end{centering}
\end{figure}

The two-dimensional posteriors in Fig.~\ref{fig6} highlight several parameter degeneracies that are less apparent from the marginalized constraints alone. In particular, $H_0$ is strongly anticorrelated with $\Omega_{\rm m}$ and positively correlated with $\Omega_\phi$, while the nearly perfect anticorrelation between $\Omega_{\rm m}$ and $\Omega_\phi$ reflects the background energy-budget constraint. The scalar-sector parameters exhibit more pronounced non-Gaussian degeneracies, most notably between $F$ and $\phi_{\rm ini}$, whose allowed region is strongly shaped by their respective parameter boundaries. By contrast, $\log\lambda$ shows only weak correlations with most of the standard cosmological parameters and remains broadly distributed, indicating that the present data have limited power to isolate the interaction strength. The addition of DESI and the low-redshift datasets noticeably reduces the geometrical degeneracies among the background parameters, although some of the degeneracy within the scalar sector persists.

\begin{figure}[t]
\centering
\includegraphics[width=0.95\linewidth]{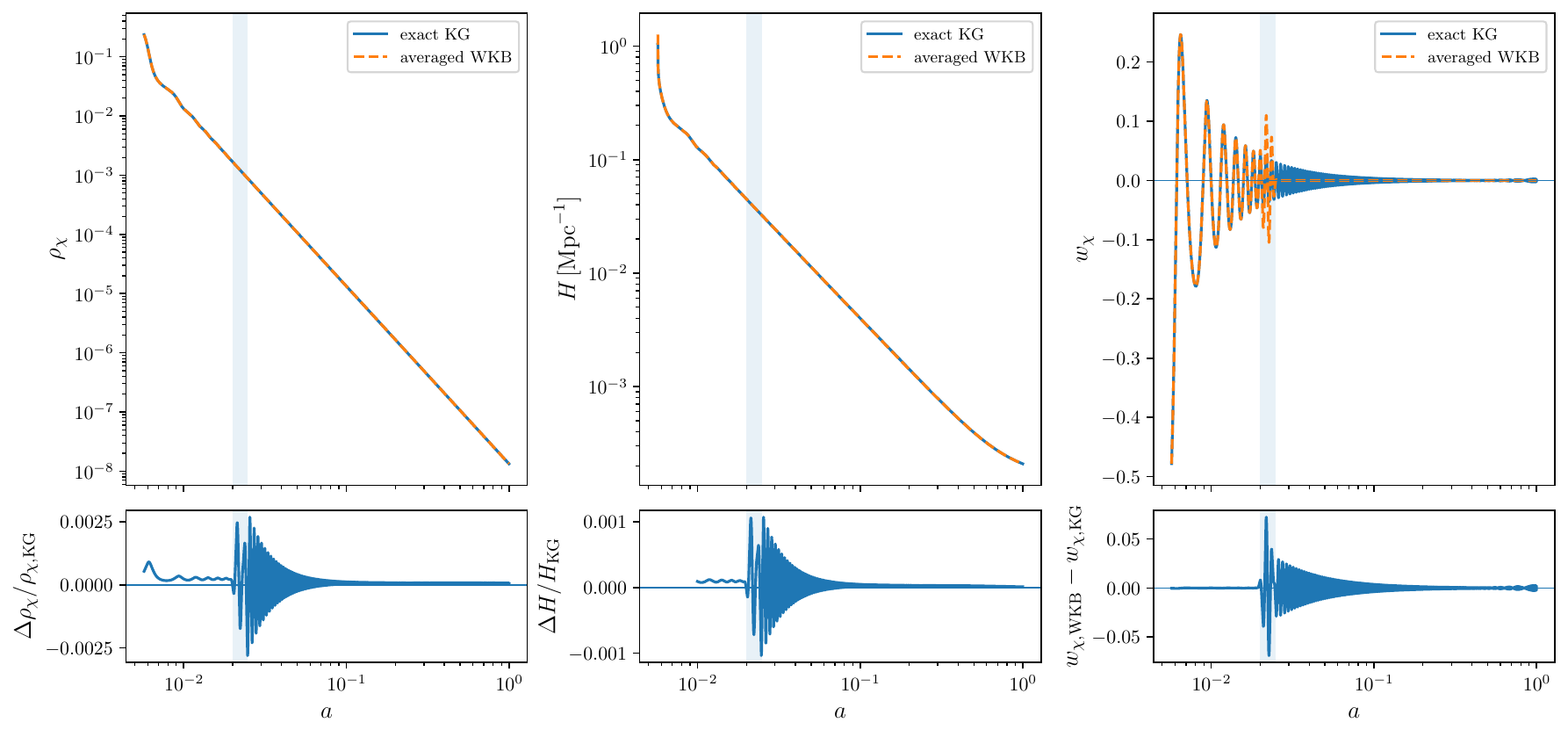}
\caption{Background comparison between the phase-resolved exact KG solution and the automatic exact-to-WKB treatment.  The shaded region marks the transition interval.  The lower panels show the corresponding residuals relative to the exact-KG solution.}
\label{fig:kg_background}
\end{figure}

\section{Numerical validation of the averaging technique}
\label{sec:numerical_validation}

In this section, we test the validity of the automatic detector and that of the averaging procedure used when the field enters the rapidly oscillating regime. The validation is carried out in complementary ways: direct comparison with an exact solution to the KG equation, variation of the detector settings, total dark-sector energy-momentum conservation through the transition, and an explicit check of the WKB conditions at the DM mass used in the MCMC analysis. The latter becomes important for non-negligible interaction which results in an effective DM mass that depends on the evolving quintessence field.

\begin{figure}[t]
\centering
\includegraphics[width=0.98\linewidth]{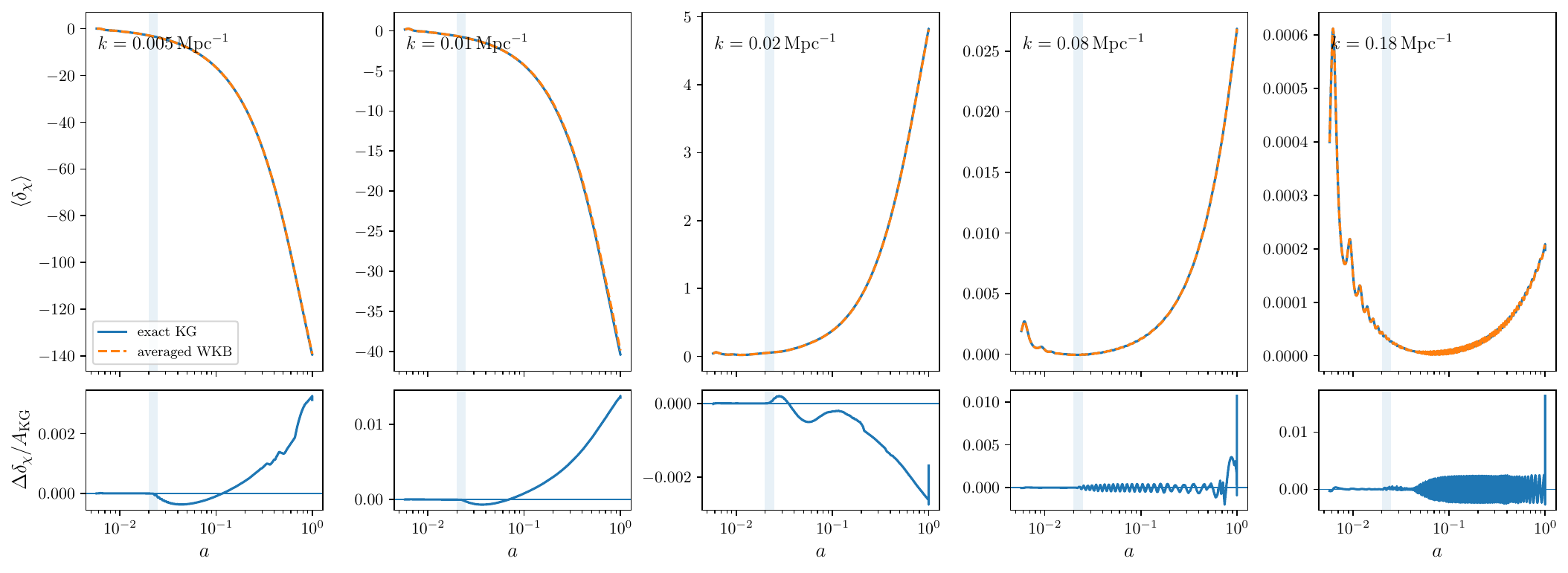}\\[1mm]
\includegraphics[width=0.98\linewidth]{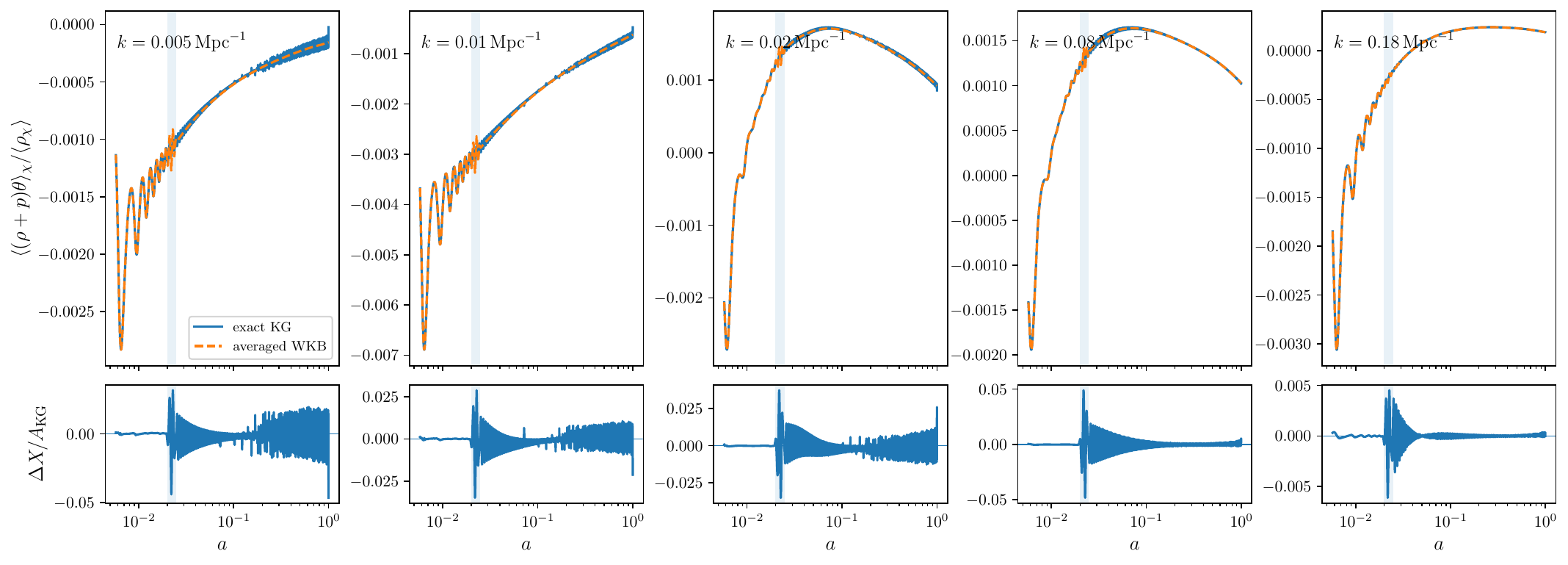}
\caption{Comparison of the cycle-averaged exact-KG perturbations with the automatic exact-to-WKB calculation for representative Fourier modes.  Top: density contrast.  Bottom: momentum density.  The residuals are normalized to the characteristic exact-KG amplitude for each mode.}
\label{fig:kg_perturbations}
\end{figure}

\subsection{Comparison with the exact KG solution}
\label{subsec:kg_validation}

In order to carry out a direct comparison between the exact KG solution and the averaged one, we choose a lighter DM mass, since the one used in this analysis, i.e., $m_\chi=10^{-20}\,{\rm eV}$, makes the numerical integration of the KG equation intractable. Therefore, we choose $m_\chi=10^{-29}\,{\rm eV}$ to test the averaging algorithm during the transition and how the exact oscillatory behavior maps to the averaged one, especially at late times.

Fig.~\ref{fig:kg_background} compares the background evolution obtained from the exact KG system (solid line) and from the automatic averaged (WKB) calculation. The two solutions in the DM energy density, $\rho_\chi$, and the Hubble parameter $H$, completely overlap and are indistinguishable by eye. The fractional deviations in both quantities are very small and become more pronounced during transition while remaining at the sub-percent level. This slight increase does not persist and decays quickly after the transition. The equation of state $w_\chi$ shows the usual rapid oscillations for the exact solution and the correct averaged solution in the WKB approximation, with deviations kept under control.

The perturbation comparison is shown in Fig.~\ref{fig:kg_perturbations} for five representative modes, $k=0.005$, $0.01$, $0.02$, $0.08$, and $0.18\,{\rm Mpc}^{-1}$.  We compare the density contrast $\delta_\chi$ and momentum density $q_\chi\equiv (\rho+p)_\chi\theta_\chi$ from the exact KG solution with the automatic (averaged) solution. The density perturbations from both solutions track each other closely over the full range of scales, with the largest deviations appearing after the transition for the lowest-$k$ modes. The momentum variable shows an oscillatory behavior and the two solutions follow each other over the entire time range with the largest deviation of 5\% observed during the transition. 

\begin{figure}[t]
\centering
\includegraphics[width=0.59\linewidth]{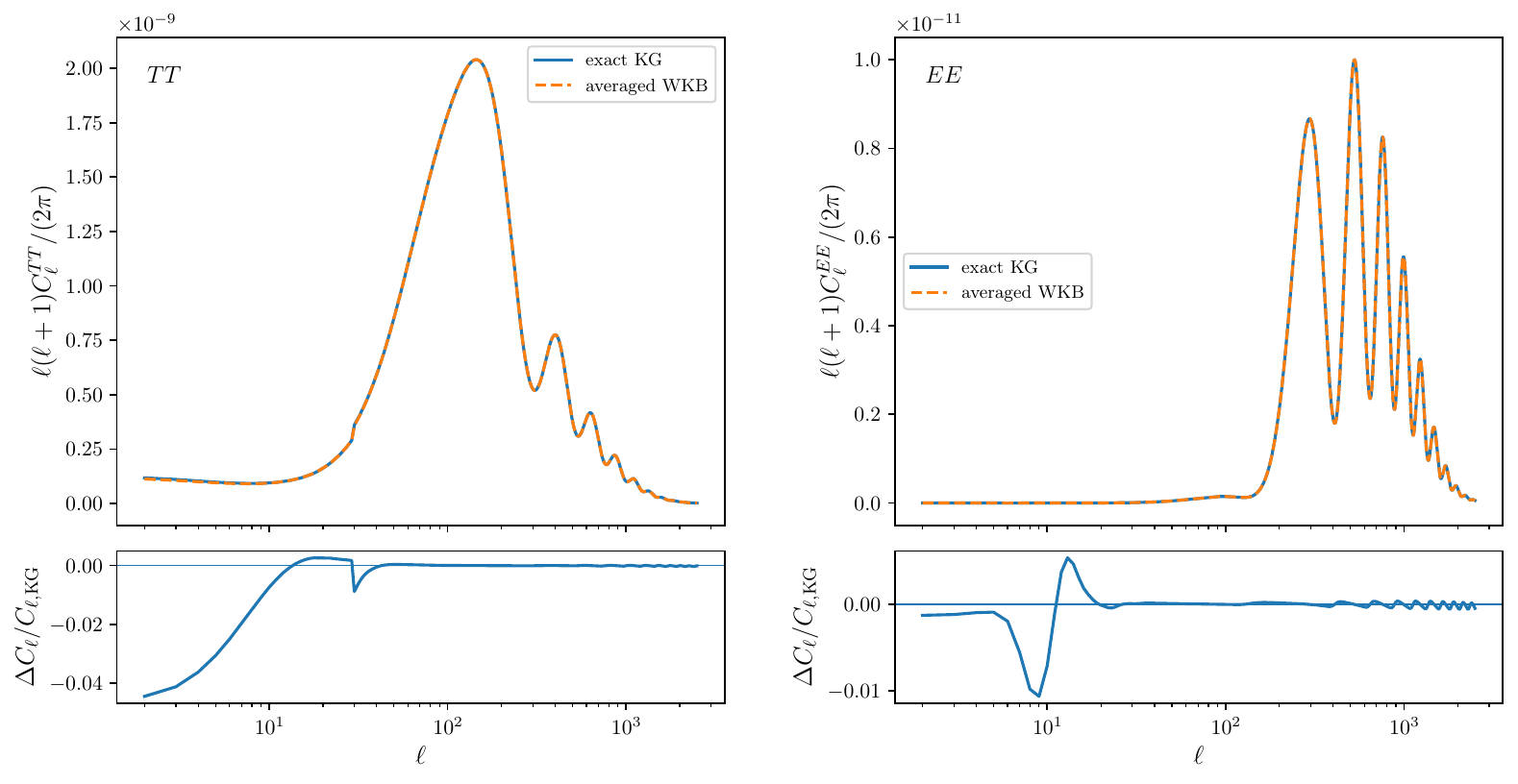}\hfill
\includegraphics[width=0.375\linewidth]{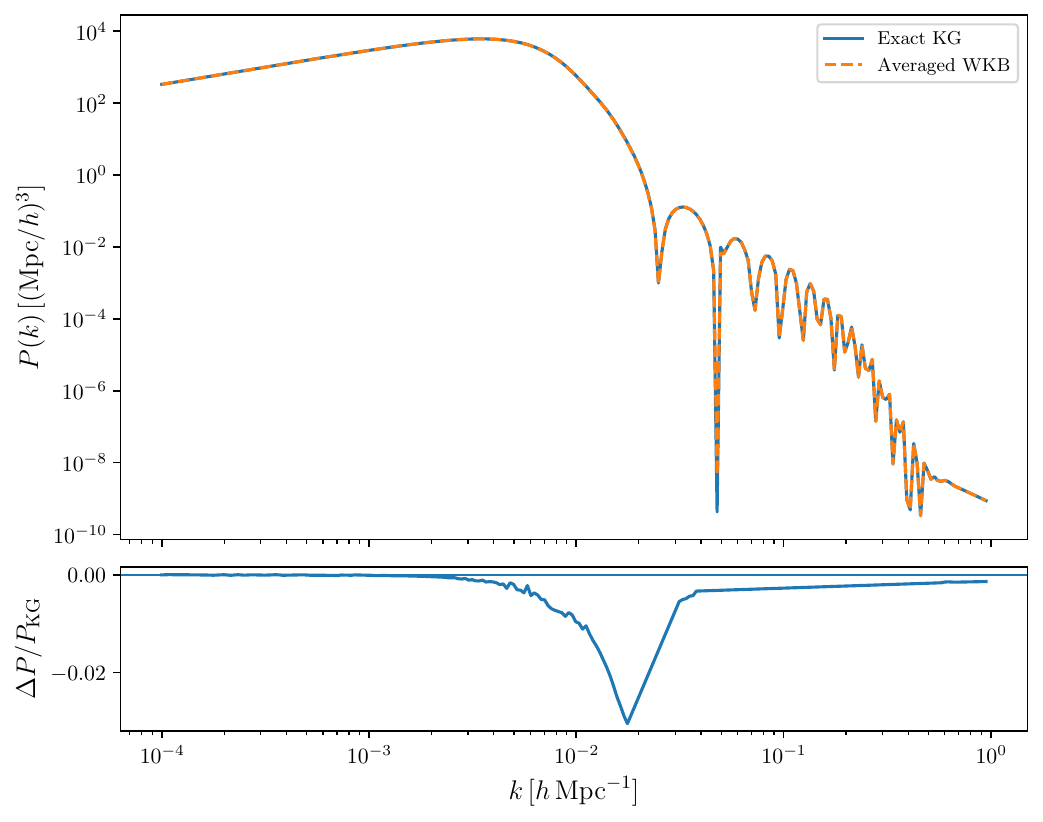}
\caption{Observable-level comparison between the exact-KG and automatic exact-to-WKB calculations.  Left: unlensed TT and EE spectra and their fractional residuals.  Right: matter power spectrum at $z=0$ and its fractional residual.}
\label{fig:kg_observables}
\end{figure}

The effect on the final observables is displayed in Fig.~\ref{fig:kg_observables}, where one can clearly see that the exact-KG and averaged spectra overlap closely.  The largest relative differences occur at the lowest CMB multipoles, where the TT residual reaches a few percent and the EE residual is at the percent level, while the agreement improves rapidly toward the acoustic region.  The matter power spectrum shows its largest localized deviation around $k\sim10^{-2}\,h\,{\rm Mpc}^{-1}$, at the few-percent level, and is substantially closer away from this region.  

These comparisons provide a clear evidence that the averaging WKB treatment used in this analysis tracks very well the exact KG solution with deviations at the percent level noticeable in the transition region. However, these deviations remain small and do not impact the main observables such as the temperature and the matter power spectra.

\subsection{Robustness of the automatic detector}
\label{subsec:detector_robustness}

The implementation of the automatic detector introduces additional control parameters as shown in table~\ref{tab1}. In this section, we perform tests on the sensitivity of the numerical results on the choice of the values of these six detector controls by varying them around their fiducial values. We, additionally, perform a 50-point Latin-hypercube scan over the full six-dimensional detector domain. We report in table~\ref{tab4} the deviations in a number of important observables as a result of variations of the detector control parameters. Varied over the range 0.5$-$0.9, the change in the ratio-drop threshold $f_R$ has no effect on the observables, while variations of $w_{\min}$ shift the transition only weakly. The parameters $\theta_{\min}$, $N_{\rm sign}$, the transition width, and the padding have a larger effect on the time the detector is triggered, but the effects on the perturbations and the associated observables (such as the CMB TT and EE power spectra and matter power spectrum $P(k)$) remain subdued since these only are predominant during the transition interval.   

For the joint scan, the largest post-transition deviations are $1.06\%$ in the amplitude-normalized density perturbation and $0.77\%$ in the momentum perturbation $q_\chi$. The median of the maximum observable deviations over the joint scan is $0.36\%$ for TT, $0.093\%$ for EE, and $0.59\%$ for $P(k)$.

\begin{table}[t]
\centering
\setlength{\tabcolsep}{3.6pt}
{\tabulinesep=1.2mm
\resizebox{\textwidth}{!}{\begin{tabular}{lcccccccc}
\toprule
Parameter & $|\Delta\ln a_{\rm mid}|$ & $\max|\Delta H/H|$ & $\max|\Delta\rho_\chi/\rho_\chi|$ & $\max|\Delta\delta_\chi|/A_{\rm fid}$ & $\max|\Delta q_\chi|/A_{\rm fid}$ & TT & EE & $P(k)$ \\
 &  & \multicolumn{7}{c}{maximum difference [\%]} \\
\midrule
$f_R$ & 0.00 & 0.00 & 0.00 & 0.00 & 0.00 & 0.00 & 0.00 & 0.00 \\
$\theta_{\min}$ & 7.16 & 0.86 & 2.12 & 1.02 & 0.30 & 0.76 & 0.71 & 0.58 \\
$w_{\min}$ & 0.25 & 0.03 & 0.07 & 0.34 & 0.01 & 0.17 & 0.72 & 0.02 \\
$N_{\rm sign}$ & 12.72 & 1.43 & 3.48 & 1.15 & 0.30 & 0.76 & 0.71 & 0.58 \\
$\Delta\theta_{\rm width}$ & 12.17 & 2.05 & 5.08 & 1.25 & 0.52 & 5.52 & 0.73 & 0.91 \\
$\Delta\theta_{\rm pad}$ & 12.05 & 1.37 & 3.32 & 0.99 & 0.34 & 0.86 & 0.12 & 0.81 \\
Joint scan & 31.14 & 3.49 & 8.82 & 1.06 & 0.77 & 5.30 & 0.71 & 2.07 \\
\bottomrule
\end{tabular}}}
\caption{Robustness of the automatic oscillation detector under variations of its numerical controls.  The entries give the largest deviations among successful automatic detections relative to the fiducial settings. Perturbation differences are evaluated after both solutions have completed the transition and evolved for one additional complete scalar oscillation. }
\label{tab4}
\end{table}

\begin{figure}[t]
\centering
\includegraphics[width=0.92\linewidth]{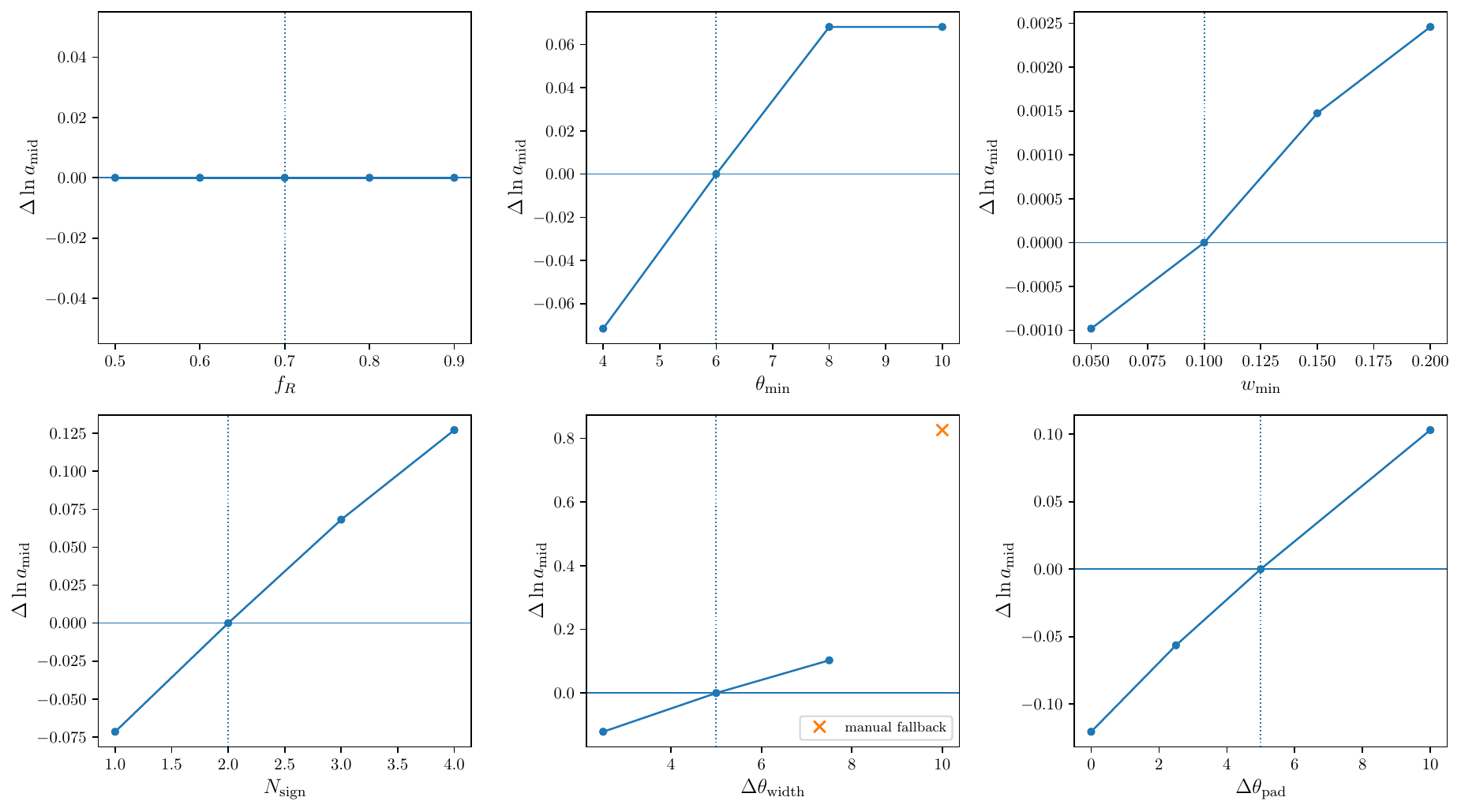}
\caption{Shift of the transition midpoint under one-at-a-time variations of the detector controls.  The vertical dotted lines denote the fiducial choices.  The marked point indicates activation of the conservative manual fallback.}
\label{fig:detector_sensitivity}
\end{figure}

\begin{figure}[t]
\centering
\includegraphics[width=0.92\linewidth]{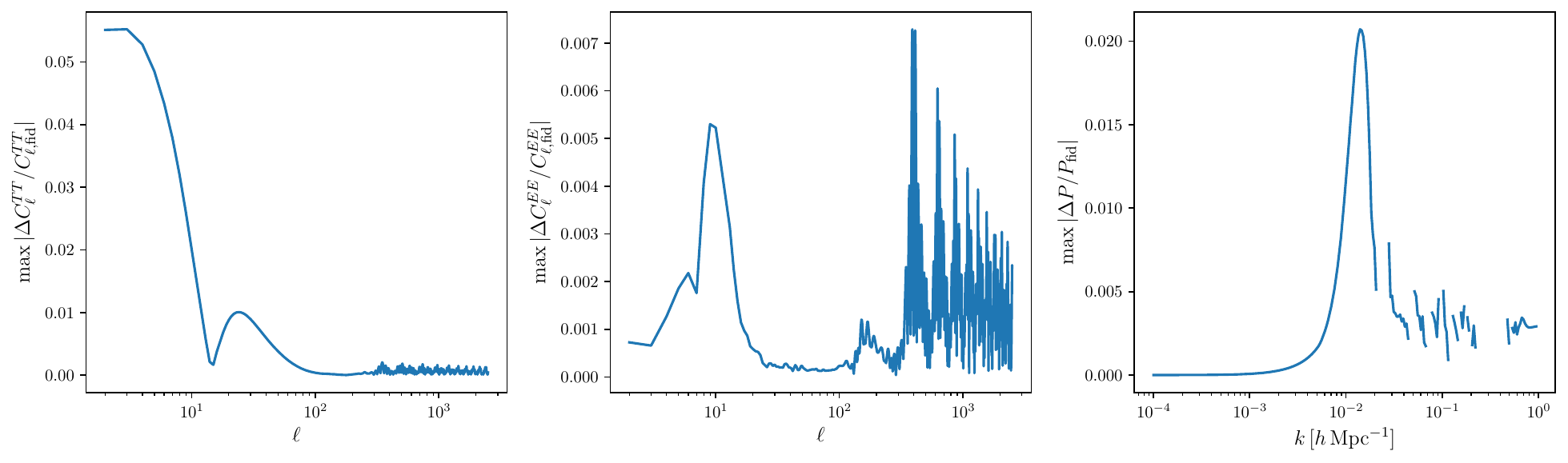}
\caption{Observable differences generated by the detector-parameter variations for $C_\ell^{TT}$, $C_\ell^{EE}$ and $P(k)$.  The curves show the maximum deviation from the fiducial automatic calculation at each multipole or wavenumber.}
\label{fig:detector_observables}
\end{figure}

Fig.~\ref{fig:detector_sensitivity} shows explicitly how each control parameter shifts the transition midpoint. The dependence is smooth and despite the percentage change in the transition scale reaching, in some cases, $\mathcal{O}(10\%)$ (which triggers the manual fallback), the effect on observables like $H$ and the density/momentum perturbations remains tiny. In Fig.~\ref{fig:detector_observables}, the relative change in the observables $C_\ell^{TT}$, $C_\ell^{EE}$ and $P(k)$ confirms that the final spectra are less sensitive than the transition time itself and the results remain robust. A separate test at the representative interacting point $\log\lambda=2.2$ gives the same qualitative conclusion: even when the maximum TT excursion reaches several percent for the most aggressive width or padding variations, the corresponding RMS differences remain below about $0.4\%$, while the EE and $P(k)$ RMS differences remain at or below the few-$10^{-3}$ level.

\subsection{Energy-momentum conservation through the transition}
\label{subsec:conservation}

\begin{figure}[H]
\centering
\includegraphics[width=0.75\linewidth]{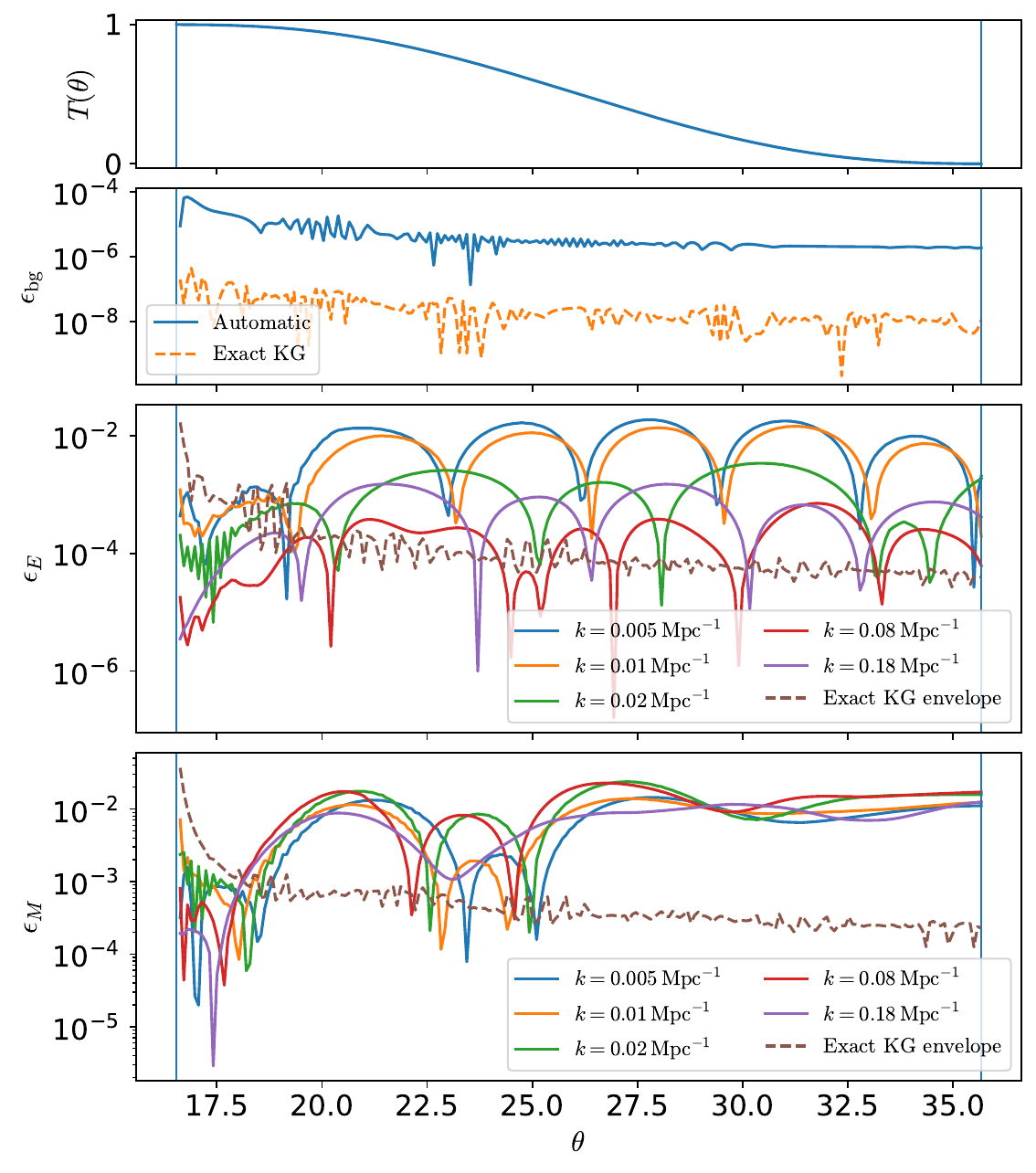}
\caption{Cumulative total-dark-sector conservation through the smooth transition from exact to averaged solutions.  From top to bottom: transition function, background and perturbation residuals with the dashed curves showing the exact-KG numerical control for the same physical interval.}
\label{fig:cumulative_conservation}
\end{figure}

The transition from the exact KG solution to the averaged solution must maintain energy-momentum conservation. So, it is imperative to check conservation of the \emph{total} interacting dark sector.  We define the total energy density and pressure of the dark sector to be $\rho_D=\rho_\chi+\rho_\phi-V_{\rm int}$ and $p_D=p_\chi+p_\phi+V_{\rm int}$, respectively, and the momentum density of the dark sector as $q_D\equiv(\rho_D+p_D)\theta_D=(\rho_\chi+p_\chi)\theta_\chi+(\rho_\phi+p_\phi)\theta_\phi$. At the background level we use
\begin{equation}
R_{\rm bg}=\frac{\text{d}\rho_D}{\text{d} N}+3(\rho_D+p_D),\qquad N\equiv\ln a,
\end{equation}
with the normalized residual
\begin{equation}
\epsilon_{\rm bg}=
\frac{|R_{\rm bg}|}
{\left|\text{d}\rho_D/\text{d} N\right|+3|\rho_D+p_D|}.
\end{equation}
In the synchronous gauge, the evolution of the density and momentum perturbations are
\begin{align}
R_E &= \delta\rho_D'
+3{\cal H}(\delta\rho_D+\delta p_D)
+q_D
+(\rho_D+p_D)\frac{h'}{2},\\
R_M &= q_D'+4{\cal H}q_D-k^2\delta p_D,
\end{align}
with primes as usual denoting derivatives with respect to conformal time and where $\delta\rho_D=\delta\rho_\chi+\delta\rho_\phi-\delta V_{\rm int}$ and $\delta p_D=\delta p_\chi+\delta p_\phi+\delta V_{\rm int}$. The normalized residuals are
\begin{align}
\epsilon_E &=
\frac{|R_E|}
{|\delta\rho_D'|+3{\cal H}|\delta\rho_D+\delta p_D|+|q_D|+
\left|(\rho_D+p_D)h'/2\right|},\\
\epsilon_M &=
\frac{|R_M|}
{|q_D'|+4{\cal H}|q_D|+k^2|\delta p_D|}.
\end{align}
The interaction energy is counted once in constructing $\rho_D$, $p_D$, and in the perturbed stress-energy tensor. To reduce sensitivity to numerical differentiation of rapidly varying quantities, Fig.~\ref{fig:cumulative_conservation} shows the cumulative, integrated versions of these residuals across the same physical transition interval.  The exact-KG solution (dashed curve) is evaluated over the identical interval and serves as a numerical control for comparison.  The background cumulative residual of the automatic solution remains at the $10^{-6}$ level through most of the transition, compared with its exact-KG counterpart near $10^{-8}$.  For the five perturbation modes, the cumulative energy and momentum residuals vary within the range $10^{-4}-10^{-2}$, with the largest values occurring for the lowest-$k$ modes.  Importantly, neither residual develops an order-unity spike or dangerous growth as the transition function changes from the exact to the averaged regime.  Therefore, the residuals from the automatic solution are greater than those from the exact KG solution, but the violation remains bounded at the percent level through the transition. 

\subsection{Fixed dark matter mass and WKB validity}
\label{subsec:mass_wkb}

In our MCMC analysis, we select a specific DM mass of $m_\chi=10^{-20}$ eV and then use Bayesian inference to extract and constrain the cosmological parameters based on this mass choice. Therefore, we do not aim to extract an independent constraint on $m_\chi$. However, we checked thoroughly that the numerical conclusions are not affected by increasing the mass above the adopted benchmark.

\begin{figure}[t]
\centering
\includegraphics[width=1.0\linewidth]{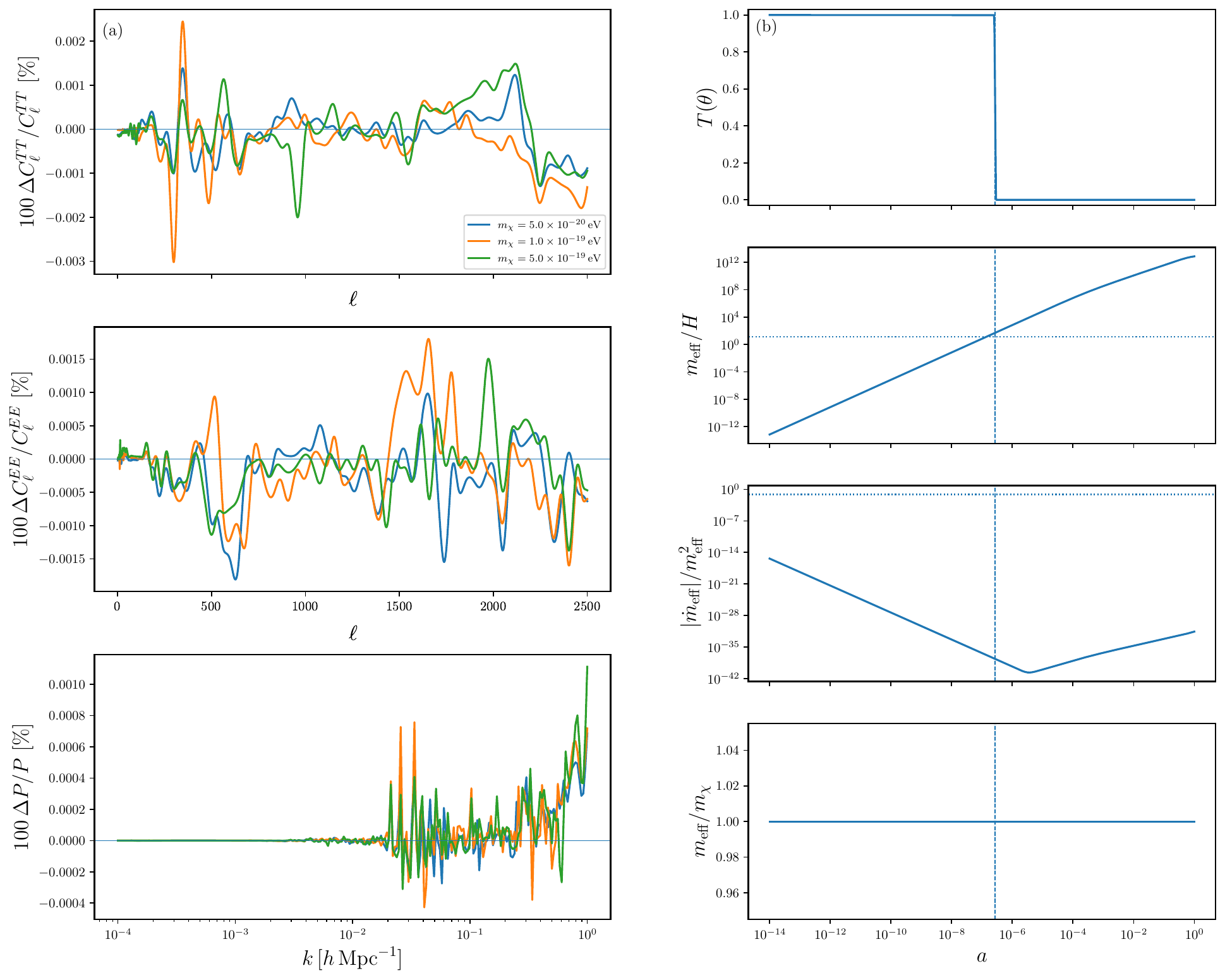}
\caption{Left panel: the fractional changes in TT, EE, and $P(k)$ for three values of the bare DM mass. Right panel (top to bottom for $m_\chi=10^{-20}\,{\rm eV}$): the transition function $T(\theta)$, the exact-branch weight, $m_{\rm eff}/H$, the effective-mass adiabaticity $|\dot m_{\rm eff}|/m_{\rm eff}^2$, and $m_{\rm eff}/m_\chi$. The vertical line marks the transition midpoint.}
\label{fig:wkb_validity}
\end{figure}

Because of the interaction, the relevant effective DM mass is
\begin{equation}
m_{\rm eff}^2=m_\chi^2+\lambda \phi^2.
\end{equation}
The correct implementation of the WKB (averaging) technique necessitates that the evolution of the expansion and the effective mass be slow compared with the oscillation frequency. To ensure this, we monitor the quantities
\begin{equation}
\epsilon_H\equiv\frac{H}{m_{\rm eff}},
\quad \text{and}\quad 
\epsilon_m\equiv\frac{|\dot m_{\rm eff}|}{m_{\rm eff}^2},
\end{equation}
where the dot denotes a derivative with respect to cosmic time.  For the chosen DM mass, the automatic transition has a midpoint at $a_{\rm mid}=2.72\times10^{-7}$ (or $z_{\rm mid}=3.68\times10^6$). At this point we obtain $\epsilon_H=1.87\times10^{-2}$, $\epsilon_m=2.10\times10^{-38}$, $m_{\rm eff}/H=53.4$, 
and the maximum value of $\max(\epsilon_H,\epsilon_m)$ anywhere across the finite transition is $2.25\times10^{-2}$. One can see from the right panel of Fig.~\ref{fig:wkb_validity} that the mass variation due to the interaction  during the transition is negligible, since $m_{\rm eff}/m_\chi\simeq1$. Therefore, the effective mass indeed evolves adiabatically (and by a large margin) in the presence of interactions.   

\begin{figure}[t]
\centering
\includegraphics[width=0.75\linewidth]{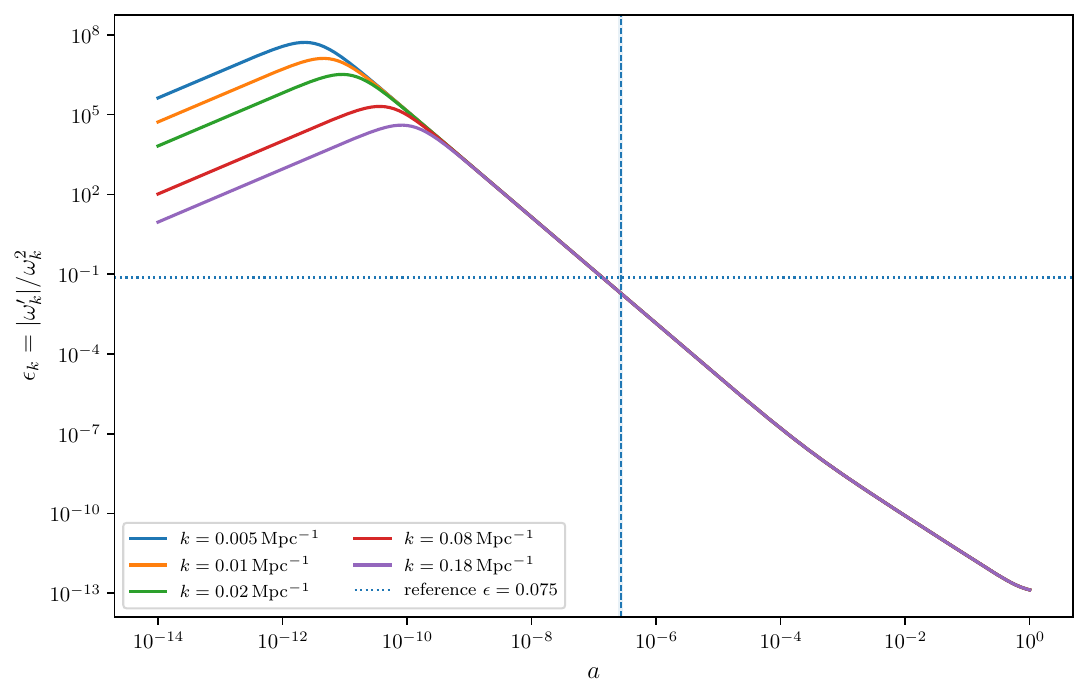}
\caption{Plot of the WKB adiabaticity parameter $\epsilon_k=|\omega_k'|/\omega_k^2$ versus the scale factor for representative Fourier modes.  The vertical line marks the transition midpoint.  The dotted horizontal line at $\epsilon=0.075$ is a conservative reference value used only to visualize the separation from the non-adiabatic regime.}
\label{fig:mode_wkb}
\end{figure}

For perturbations, the frequency of a Fourier mode in conformal time is
\begin{equation}
\omega_k^2=k^2+a^2m_{\rm eff}^2,
\end{equation}
and the corresponding WKB condition is
\begin{equation}
\epsilon_k\equiv\frac{|\omega_k'|}{\omega_k^2}\ll1,
\end{equation}
where the prime denotes a derivative with respect to conformal time.  Fig.~\ref{fig:mode_wkb} shows this quantity for the same five modes used in the perturbation validation.  At the transition point, all modes have values of $\epsilon_k\simeq1.87\times10^{-2}$, with a maximum of approximately $2.25\times10^{-2}$ across the full transition. The horizontal line at $\epsilon=0.075$ is only a conservative reference, while the relevant requirement is $\epsilon_k\ll1$.  The measured values are comfortably below unity and below this reference throughout the averaging interval. 

Finally, we test the effect of increasing the bare mass from $10^{-20}\,{\rm eV}$ to $5\times10^{-20}$, $10^{-19}$, and $5\times10^{-19}\,{\rm eV}$ on the CMB temperature power spectrum and the matter power spectrum. The left panel of Fig.~\ref{fig:wkb_validity} shows the resulting changes in $C_{\ell}^{TT}$ and $C_{\ell}^{EE}$ through $\ell=2500$ (the upper two plots) and in the matter power spectrum (lower plot). Over this range of DM masses, notice that the TT and EE spectra vary only at $\mathcal{O}(10^{-3})\%$, while the matter power spectrum changes by at most approximately $10^{-3}\%$ over the plotted range. Therefore it is clear that fixing the DM mass to $m_\chi=10^{-20}\,{\rm eV}$ does not introduce a relevant numerical sensitivity into the parameter analysis presented here.

In this validation analysis of the automatic detector, we checked that the exact KG solution and the averaged solution track each other well after the detector is switched on and that the averaging happens after the scalar field has entered an adiabatic oscillatory regime. We have shown that the final observables are not sensitive to reasonable changes in the detector's settings and that energy-momentum conservation holds during the transition. Furthermore, the choice of a fixed DM mass in MCMC lies safely within the regime where the averaged treatment is numerically stable.

Before closing, a word about the unit of the coupling $\lambda$ is in order. From particle physics (using natural units), the fields in Eq.~(\ref{vint}) are canonically normalized dimensionful fields and so $\lambda$ is dimensionless. However, in \code{CLASS}, the fields are in units of $M_{\rm Pl}$, while for canonically normalized dimensionful fields $X$ and $\varphi$, the interaction is $g X^2\varphi^2/2$, where $g$ is a dimensionless coupling. In \code{CLASS}, the fields are rescaled as $\chi=X/M_{\rm Pl}$ and $\phi=\varphi/M_{\rm Pl}$, so the interaction potential becomes $gM_{\rm Pl}^4 \chi^2\varphi^2/2$. But since the potential in CLASS has units of $M_{\rm Pl}^2$ (since $V_{\rm CLASS}=V_{\rm Physical}/M_{\rm Pl}^2$), then $\lambda\equiv gM_{\rm Pl}^2$. The coupling in \code{CLASS} units has dimensions of mass-squared which is quoted in Mpc$^{-2}$.

\section{Conclusion}\label{sec:conclusion}

The correct and accurate description of scalar field cosmology requires methods to treat fast oscillations which makes finding a solution to the Klein-Gordon equation numerically intractable. The existing literature focuses on two methods: (1) solving the exact Klein-Gordon equation prior to the onset of fast oscillations and then switching to solving the averaged (fluid) equations once these oscillations begin, or (2) introducing new variables that allows us to rewrite the Klein-Gordon equation in a way which makes it easier to handle numerically. Both of these approaches rely on an estimate of when the rapid oscillations in the scalar field begin. An incorrect estimate can either skew the results and allow for inaccuracies to propagate into the calculations of observables or can slow down numerical codes which can be problematic especially when running MCMC simulations.

In this work, we presented an accurate method, which we implemented in \code{CLASS}, to automatically detect the onset of oscillations and trigger the averaging process. The algorithm also employs a second independent method to verify the oscillations are indeed happening by checking for sign changes in the DM equation of state. We demonstrated the efficiency and accuracy of this technique by applying it to the field-theoretic model of cosmology, QCDM, which includes a DM-DE interaction term. The modified \code{CLASS} code with the detection and averaging techniques is made public on the \href{https://github.com/aabouibrahim/class_field_theoretic}{Github} page. The \code{CLASS} model is run with the MCMC tool \code{Cobaya} to extract an updated set of constraints on the model parameter space of QCDM. The full data set results in tighter constraints on the cosmological parameters with a preference for higher $H_0$, while the interaction term remains weakly constrained. Using the Bayes factor, we have seen that the data provides weak evidence in favoring $\Lambda$CDM over QCDM, despite it having a larger parameter space compared to $\Lambda$CDM.

\vspace{1cm}

{\bf Acknowledgments:} 
The research of PN was supported in part by the NSF Grant PHY-2209903, while the work of AA is supported in part by the NASA CT Space Grant Prosum number P-2590. This work used the Explorer cluster at Northeastern University for MCMC simulations.

\end{document}